\documentclass[12pt]{article}
\usepackage{amssymb,amsmath}
\usepackage[english]{babel}
\usepackage[round,authoryear]{natbib} 
\usepackage[]{graphicx}               
\begin{document}
%
%
%
\title{    Resonant energy exchange in nonlinear oscillatory chains \\%
    and Limiting Phase Trajectories: from small to large systems.}
%
%
\author{
Leonid I. Manevitch\textsuperscript{*} and Valeri V. Smirnov\textsuperscript{\dag}\\
Institute of Chemical Physics, RAS \\
4 Kosygin str., PO 119991, Moscow, Russia \\
\textit{e-mail:}~\textsuperscript{*}\textit{lmanev@chph.ras.ru},~\textsuperscript{\dag}\textit{vvs@polymer.chph.ras.ru} \\
}%
%
%
    \maketitle
%

    \begin{abstract}
We present an adequate analytical approach to the description of nonlinear vibration with strong energy exchange between weakly coupled oscillators and oscillatory chains. The fundamental notion of the limiting phase trajectory (LPT) corresponding to complete energy exchange is introduced. In certain sense this is an alternative to the nonlinear normal mode (NNM) characterized by complete energy conservation. Well-known approximations based on NNMs turn out to be valid for the case of weak energy exchange, and the proposed approach can be used for the description of nonlinear processes with strong energy exchange between weakly coupled oscillators or oscillatory chains. Such a description is formally similar to that of a vibro-impact process and can be considered as starting approximation when dealing with other processes with intensive energy transfer. At first we propose a simple analytical description of vibrations of nonlinear oscillators. We show that two dynamical transitions occur in the system. First of them corresponds to the bifurcation of anti-phase vibrations of oscillators. And the second one is caused by coincidence of LPT with separatrix dividing two stable stationary states and leads to qualitative change in both phase and temporal behavior of the LPT (in particular, temporal  dependence of the amplitude becomes resembling that for vibro-impact vibrations). 
Next problem under consideration relates to intensive intermodal exchange in the periodic nonlinear systems with finite (n$>$2) number of degrees of freedom. We consider two limiting cases. If the number of particles is not large enough, the energy exchange between nonlinear normal modes in two-dimensional integral manifolds is considered. When the number of the particles increases the energy exchange between neighbor integral manifolds becomes important that leads to formation of the localized excitations resembling the breathers in the one-dimensional continuum media.
    \end{abstract}

%

\subsection*{Introduction}

Understanding the mechanisms of energy transfer and localization is one of the key problems in physics, mechanics, chemistry and biology. This problem is strongly connected with searching the elementary excitations preserving their characteristics under mutual interactions. The nonlinear normal mode (NNM) that is an extension of well-known normal mode in the linear systems (LNMs) is the important type of elementary excitation. Contrary to LNMs, the NNMs do not possess the superposition property. Besides, the NNMs can demonstrate a spatial localization and their number may exceed the number of degrees of freedom ~[\cite{book1, book2}]. In the linear systems the energy cannot be transferred by LNM (because they have homogeneous structure and infinite prolongation). As for some combination of LNMs, in particular, the wave packets, they can transfer the energy, but inevitably decay because of dispersion phenomenon ~[\cite{book3}]. If nonlinearity is taken into account, the solitonic mechanism of energy transfer (by non-topological or topological solitons as well as by envelope solitons (breathers)) becomes possible ~[\cite{book4}]. But the conventional results in this field relate to infinite continuum systems or oscillatory chains. In this connection two important questions arise: (1) what is a mechanism of energy transfer in relatively small periodic oscillatory chains? And (2) how does transition to localized excitations proceeds when the number of the particles in periodic chain increases?

The object of the paper is to give the answers on these questions. At that is shown that the conception of Limiting Phase Trajectory (LPT) has a key role in the understanding the regularities of intensive energy transfer in the oscillatory chains. This conception was first introduced in the paper ~[\cite{book5}] in the application to two degrees of freedom (2DoF) system. We consider first the intensive interaction of degenerate NNMs (which may be correlated with q-breathers in the Fourier space of the system) in the invariant manifolds, described by LPTs, as mechanism of slow energy transfer by periodic traveling (envelope) waves. Then it is shown that a closeness of the frequencies, corresponding to different integral manifolds, provides formation of stationary or mobile localized excitations resembling the breather in the infinite chain when the number of the particles increases.

     
\section{Two weakly coupled nonlinear oscillators}

\begin{figure}[htbp]
            \centering
        \includegraphics [width=69.5mm, height=21.2mm] {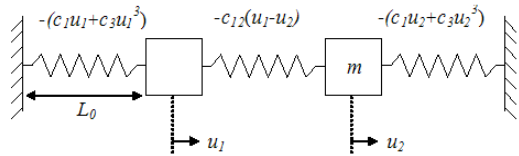}
        \caption{The system under consideration}
	\label{f1}
    \end{figure}

Let us consider first the simplest nonlinear problem of energy transfer in two weakly coupled nonlinear oscillators with cubic restoring forces (Fig.\ref{f1}).


 This problem can be described by the following system of two nonlinear equations (in dimensionless form):

\begin{equation} \label{eq1} \begin{array}{l} {\frac{d^{2} U_{1} }{d\tau _{1}^{2} } +U_{1} +2\beta \varepsilon \left(U_{1} -U_{2} \right)+8\alpha \varepsilon \, U_{1}^{3} =0,\; 2\beta \varepsilon =\frac{c_{12} }{c_{1} } ,} \\ {\frac{d^{2} U_{2} }{d\tau _{1}^{2} } +U_{2} +2\beta \varepsilon \left(U_{2} -U_{1} \right)+8\alpha \varepsilon \, U_{2}^{3} =0} \end{array} \end{equation} 

where 

\begin{equation} \label{eq2} U_{j}^{} =\frac{u_{j0}^{} }{L_{0}^{} } ,\; \tau _{1} =\sqrt{\frac{c_{1} }{m} } ,\; 8\alpha \varepsilon =\frac{c_{3}^{} L_{0}^{2} }{c_{1} } , \end{equation} 

\textit{$c_1$} and \textit{$c_3$} are the linear and nonlinear stiffnesses of the first and second oscillators, respectively, and $c_{12}$ is the stiffness of the coupling spring. Introducing the complex variables:

\begin{equation} \label{eq3} \begin{array}{l} {\varphi _{1} =e^{-i\tau _{1} } \left(\frac{dU_{1} }{d\tau _{1} } +iU_{1} \right),\; \; \varphi _{1}^{*} =e^{i\tau _{1} } \left(\frac{dU_{1} }{d\tau _{1} } -iU_{1} \right),} \\ {\varphi _{2} =e^{-i\tau _{1} } \left(\frac{dU_{2} }{d\tau _{1} } +iU_{2} \right),\; \; \varphi _{2}^{*} =e^{i\tau _{1} } \left(\frac{dU_{2} }{d\tau _{1} } -iU_{2} \right)} \end{array} \end{equation} 

and slow time $\tau_2=\varepsilon\tau_1$ (along with the fast time $\tau_1$), one can use the following two-scale expansions

\begin{equation} \label{eq4} \varphi _{j} \left(\tau _{1} ,\tau _{2} \right)=\sum _{n}\varphi _{j,n} \left(\tau _{1} ,\tau _{2} \right) \, \varepsilon ^{n} ,\; \; j=1,2. \end{equation} 

After corresponding calculations taking into account Eqs. (\ref{eq1}-\ref{eq4}) we arrive at the equations of the principal asymptotic approach

\begin{equation} \label{eq5} \begin{array}{l} {\frac{df_{1} }{d\tau _{2} } +i\, \beta \, f_{2} -3i\alpha \left|f_{1}^{2} \right|f_{1} =0,} \\ {\frac{df_{2} }{d\tau _{2} } +i\, \beta \, f_{1} -3i\alpha \left|f_{2}^{2} \right|f_{2} =0,} \\ {\varphi _{j} =e^{i\beta \tau _{2} } f_{j,\; \; } j=1,\; 2} \end{array} \end{equation} 

which describe, as well as the mechanical system discussed, others including optic couplers ~[\cite{book6}]. This system (Fig. \ref{f1}) is fully integrable and has two integrals:

\begin{equation} \label{eq6} H=\beta \left(f_{2} f_{1}^{*} +f_{1} f_{2}^{*} \right)-\frac{3}{2} \alpha \left(\left|f_{1} \right|^{4} +\left|f_{2} \right|^{4} \right)\, ,  \end{equation} 

\begin{equation} \label{eq7} N=\left|f_{1} \right|^{2} +\left|f_{2} \right|^{2} \,  \end{equation} 

The best way to address this is to use Eq. \ref{eq7} and the coordinates \textit{$\theta $} and $\Delta$, where 

\begin{equation} \label{eq8} f_{1} =\sqrt{N} \cos \theta \, e^{i\delta _{1} } ,\; f_{2} =\sqrt{N} \cos \theta \, e^{i\delta _{2} } ,\; \Delta =\delta _{1} =\delta _{2} , \end{equation} 

\begin{equation} \label{eq9} \frac{d\theta }{d\tau _{2} } =\beta \sin \Delta ,\; \sin 2\theta \frac{d\Delta }{d\tau _{2} } =2\beta \cos 2\theta \cos \Delta +\frac{3}{2} \alpha \, N\sin 4\theta . \end{equation} 

The integral of Eq. \ref{eq9} has the form following from Eq. \ref{eq6}

\begin{equation} \label{eq10}
H_{0}^{} =\cos \Delta +k^{2} \sin 2\theta ,  
\end{equation}

where $k=\frac{3\alpha \, N}{4\beta } ,\; \alpha >0$

The system \ref{eq9} is strongly nonlinear even in the case of initially linear problem. 

\begin{figure}[htbp]
            \centering
        \includegraphics [width=107.6mm, height=110.2mm]{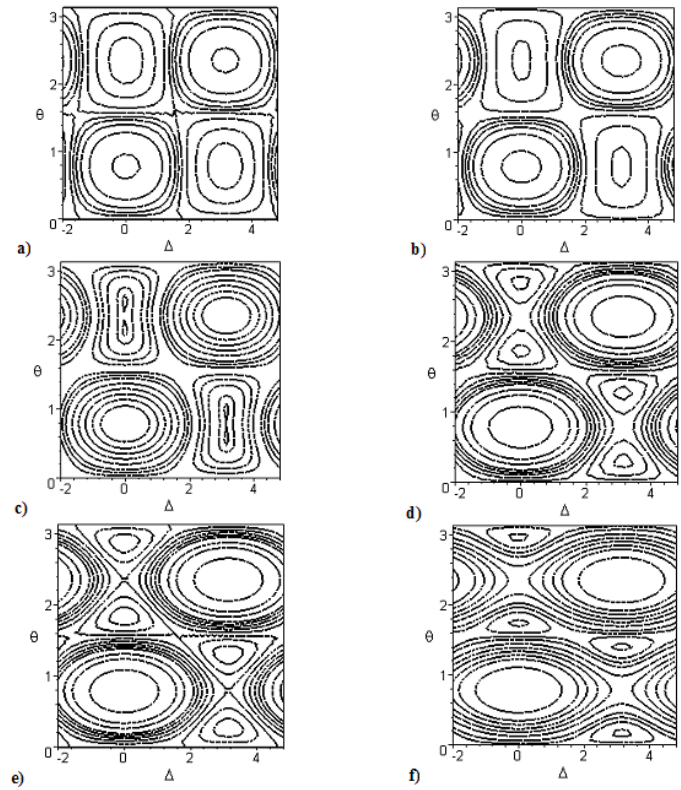}
        \caption{Phase trajectories in the $(\theta, \Delta)$ plane for: a) \textit{k }= 0.2, b) \textit{k }= 0.4, c) \textit{k }= 0.55, d) \textit{k}=0.9, e) \textit{k }= 1., f) \textit{k }= 1.5}
	\label{f2}
    \end{figure}

Let us present plots of the phase trajectories for different values of \textit{k }in Fig. \ref{f2}. (Because of the phase plane periodicity one only has to consider the two lower quadrants.)

Let us show that we can consider the LPT as fundamental solution (similarly to NNMs) whose behavior determines the second dynamic transition in the behavior of the oscillatory system. The LPT, which is far from the stationary points, can then be used as a generating solution to construct close trajectories with strong energy transfer.

The value of \textit{H }for the LPT in integral \ref{eq10} is equal to zero. Therefore the variables \textit{$\theta $} and $\Delta$ in this case are connected by the equation

\begin{equation}
 \label{eq11} \cos \Delta =-k\sin 2\theta ,  
\end{equation} 

so that $\sin \Delta =\pm \sqrt{1-k^{2} \sin ^{2} 2\theta } $. 

Then the first of the equations \ref{eq9} can be written as:

\begin{equation}
 \label{eq12} \frac{d\theta }{d\tau _{2} ^{} } =\pm \beta \sqrt{1-k^{2} \sin ^{2} 2\theta }  
\end{equation}

The solution of Eq. (\ref{eq12}) for the plus sign is the elliptic Jacobi's function $ \theta\,=\, (1/2)am(2\beta \tau_2,k)$. Because $ 0 > \theta >\pi/2 $ by definition, one can use the negative sign for $(2n-1)\pi < 2\beta \tau_2 < 2n\pi $, $ n=1, 2, 3\dots  $

Thus we arrive at the final solution

\begin{equation} \label{eq13} \theta =\frac{1}{2} \left|am\, \left(2\beta \tau _{2} ,k\right)\, \right|\, ,\; \Delta =\pm \arccos \left[k\, sn\, \left(2\beta \tau _{2} ,k\right)\right],  \end{equation} 

with period \textit{K}(\textit{k}), i.e., the complete elliptic integral of the first kind (for the in-phase oscillations). 

The solution for the out-of-plane oscillations is

\[\theta =\frac{1}{2} \left|am\, \left(2\beta \tau _{1} ,k\right)\, \right|\, ,\; \Delta =\pi \pm \arcsin \left[n\, \left(2\beta \tau _{1} ,k\right)\right].\] 

\begin{figure}[htbp]
            \centering
        \includegraphics [width=53.8mm, height=55.1mm]{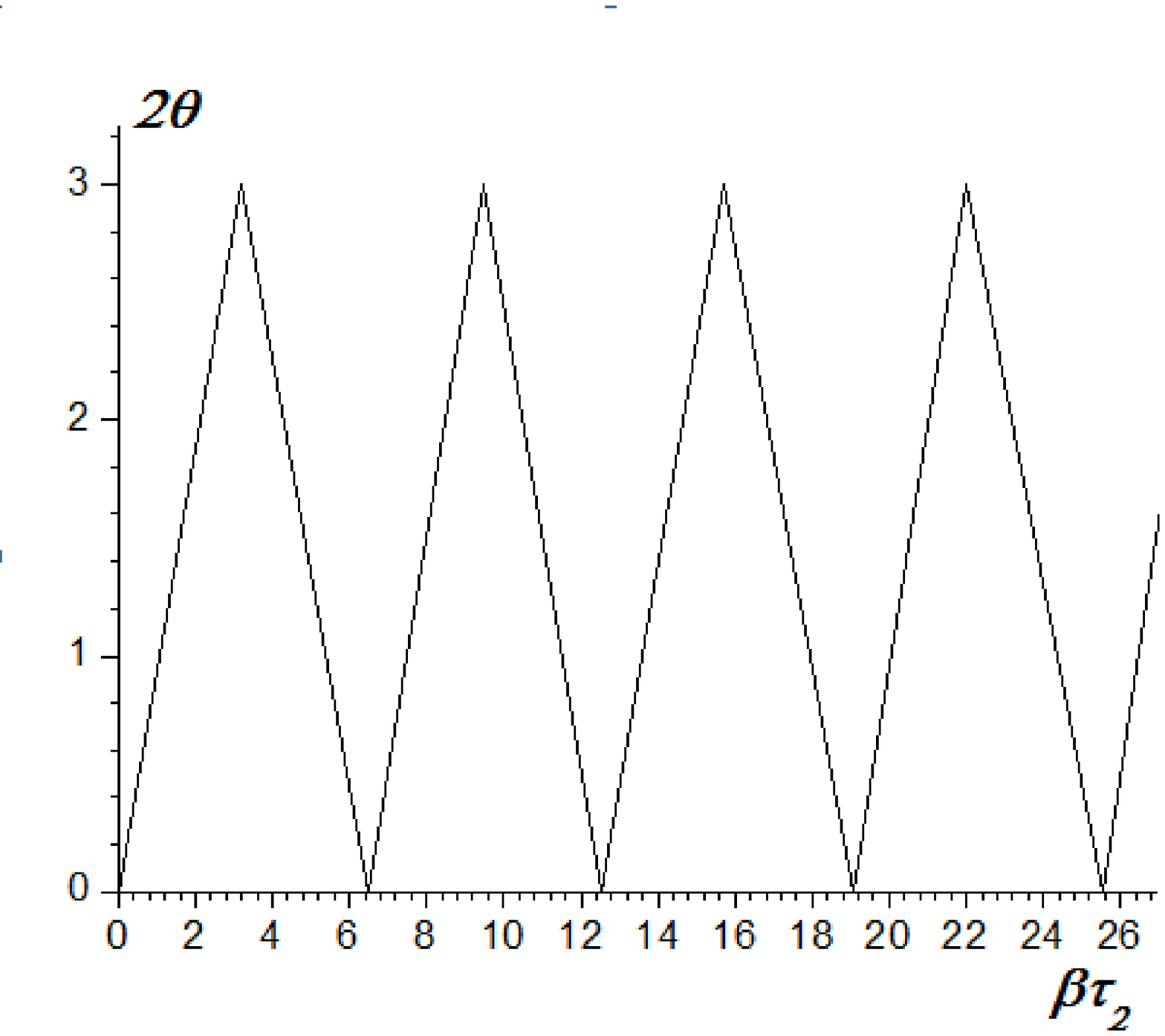}
        \caption{The function $2\theta =\left|am\left(2\tau _{2} ,k\right)\right|$ for k=0.5}
	\label{f3}
    \end{figure}

\begin{figure}[htbp]
            \centering
        \includegraphics [width=53.8mm, height=55.1mm]{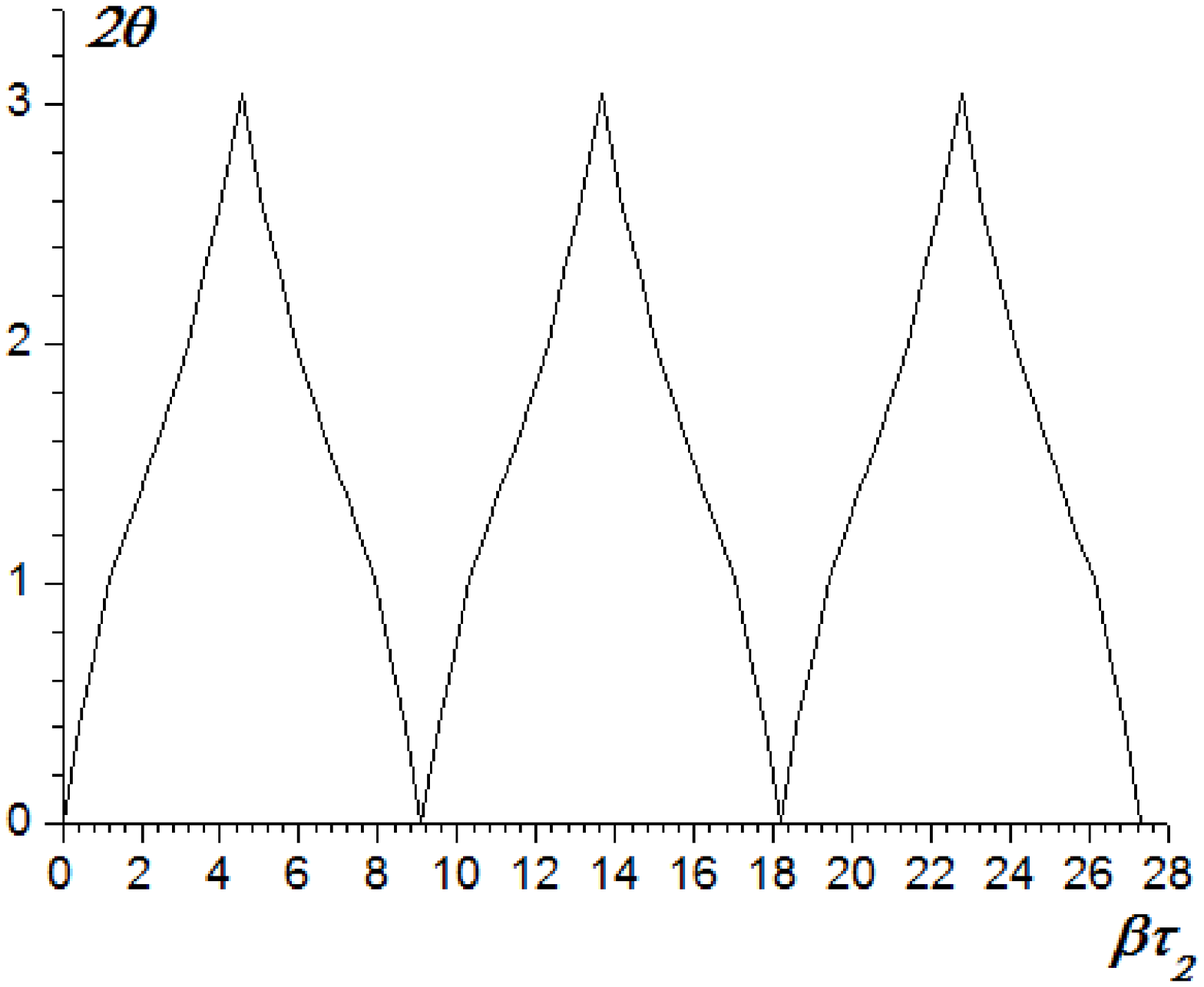}
        \caption{The function $2\theta =\left|am\left(2\tau _{2} ,k\right)\, \right|$ for k=0.9}
	\label{f4}
    \end{figure}

The periodic functions (\ref{eq13}) are not smooth; $\Delta(2\beta \tau_2)$ has breaks at the points $2 \beta\tau_2=(2n-1)\pi$, \textit{n}=0,1 ,\dots , and $\theta (2\beta \tau_2)$ has discontinuities in the derivative at these points (in terms of distributions $d\theta/d\tau_2 = (2\beta/\pi)\Delta$). Plots of $\theta(2\beta \tau_2)$ for two values of parameter \textit{k} are presented in Figs. \ref{f3} and \ref{f4}. The value \textit{k }= 0.5 corresponds exactly to the first dynamic transition. However, the solution for the LPT (Fig. \ref{f3}) is still close to that of the linear case, except for a small change of period. Only for values of \textit{k} that are close to unit the deflections from an exact saw tooth profile and the change of period become noticeable.

The second dynamic transition occurs when \textit{k }= 1. In this case one can find a simple analytical solution corresponding to the LPT:

\begin{equation}
 \label{eq14} 2\beta \tau _{2} =\int _{0}^{2\theta }\frac{d\left(2\theta \right)}{\cos 2\theta }  ,\; \; \theta =\frac{1}{2} \arcsin \frac{1-e^{-2\beta \tau _{2} } }{1+e^{-2\beta \tau _{2} } } ,
  \end{equation} 

It can be seen from Eq. \ref{eq14} that the LPT actually coincides with a separatrix: if $k \rightarrow~1$, ~$\theta \rightarrow\pi/4$ when ~$\tau_2 \rightarrow \infty$. 

\begin{figure}[htbp]
            \centering
        \includegraphics[width=100.4mm, height=28.7mm]{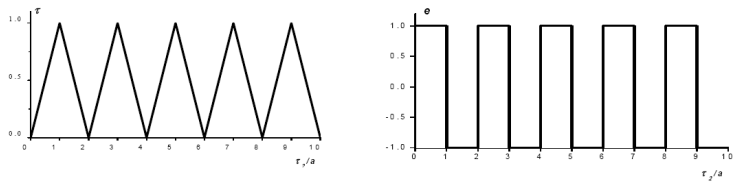}
        \caption{The non-smooth basic functions $\tau (\tau_{2}/a$), $ e  (\tau_{2}/a$), where $2a$ is a halved period (in time $\tau_{2}$)}
	\label{f5}
    \end{figure}

It is convenient to introduce two non-smooth functions $\tau(\tau_2)$, $e(\tau_2)$ (Fig. \ref{f5}). Similar functions (but with alternating signs of dependent variables) were introduced first in ~[\cite{book7}] to calculate the smooth vibrational regimes close to vibro-impact ones. The development of corresponding techniques is also presented in ~[\cite{book7}]. Here we use similar notation.

We would like to show that the natural area for application of these non-smooth basic functions is the description of beats (using the variables $\theta $ and $\Delta$) and close trajectories with strong energy transfer. Actually in the case \textit{k }= 0 (the linearized system) the solution (\ref{eq8}) can be rewritten in the form $\theta =(\pi/2)\tau$, $\Delta=(\pi/2)e$, $\tau= \tau(\tau_2/a)$, $e=e(\tau_2/a)$, where $a =\pi/2\beta $ (exactly as in a vibro-impact process with velocity $\Delta=\pi/2$). After introducing the basic functions $\tau(\tau_2/a)$, $e(\tau_2/a)$, we can present the solution as 

\begin{equation}
 \label{eq15} \theta =X_{1} (\tau )+Y_{1} (\tau )\, e\, \left(\frac{\tau _{2} }{a} \right),\; \; \Delta =X_{2}^{} (\tau )+Y_{2}^{} (\tau )\, e\left(\frac{\tau _{2} }{a} \right),  
\end{equation} 

where the smooth functions $X_i(\tau)$, $Y_i(\tau)$ satisfy Eqs.(\ref{eq9}):

\begin{equation} 
\label{eq16} 
\begin{array}{l} \frac{\partial }{\partial \tau } \left\{\begin{array}{c} {X_{1} } \\ {Y_{1} } \end{array} \right\}=\frac{1}{2} a\beta [\sin (X_{2} +Y_{2} )\mp \sin (X_{2} -Y_{2} )] \\
 \frac{\partial }{\partial \tau } \left\{\begin{array}{c} {X_{2} } \\ {Y_{2} } \end{array}\right\}=a\beta [cot 2(X_{1} +Y_{1} )\cos (X_{2} +Y_{2} ) \\ 
 \mp cot 2(X_{1} -Y_{1} )\cos (X_{2} -Y_{2} )]  +\frac{3a}{2} \alpha \, N[\cos _{} 2(X_{1} +Y_{1} )\mp \cos _{} 2(X_{1} -Y_{1} )] \end{array} 
\end{equation} 

Then, we can search for the solution of Eq. (\ref{eq16}) in the form of power expansions in the independent variable $\tau $:

\begin{equation} \label{eq17}
 X_{i}^{} =\sum _{l=0}^{\infty }X_{i,\, \, l\, }  \tau _{}^{l} ,\; \; Y_{i}^{} =\sum _{l=0}^{\infty }Y_{i,\, l\, }  \tau _{}^{l} ,\; \; i=1,\; 2 
\end{equation} 

where the generating solution is the linear beat: $X_{1,0} = 0$, $X_{1,1} =\pi/2$, $Y_{1,0} = 0$, $X_{2,0} = 0$,  $Y_{2,0} =\pi/2$, satisfying exactly the $\theta - \Delta$ equations for the case of the strongest beat. It can be proved that the presentation (\ref{eq15}), taking into account (\ref{eq17}), actually recovers the exact solution of the nonlinear problem for the most intensive energy transfer between the oscillators. As this takes place, the expansions (\ref{eq17}) restore the exact local representation of the corresponding elliptic function (near $\tau = 0$), but the expressions (\ref{eq15}) allow the prediction of the exact global behavior of the system. It is important to note that, even for large enough values of \textit{k}, the solution appears close to that of linear beats; the only change is the barely seen curvature of the lines that are straight for linear beats, and a change of the period.  

One can find corresponding corrections by considering the next order of approximations, namely, $X_{1,0} = 0$, $X_{1,1} = \alpha\beta$, $X_{1,3} = -(2/3) (\alpha\beta)^3 k^2$, $Y_{2,0} =\pi/2$, $Y_{2,1} =\pi/ 2$, $Y_{2,1} = 2\alpha \beta k$, $Y_{2,3} = 4/3(\alpha \beta )^3$, \dots , which coincide with those in the expansions of the exact solution.

\section{Resonant energy transfer and energy localization in finite FPU systems with periodic boundary conditions}

\subsection*{FPU systems with periodic boundary conditions}

Let us consider a common FPU-system ~[\cite{book8}], the Hamiltonian of which can be written as follows:

\begin{equation} \label{eq18} H_{0} =\sum _{j}^{N}\frac{1}{2} p_{j} ^{2} +\frac{1}{2} (q_{j+1} -q_{j} )^{2} +\alpha \frac{1}{3} (q_{j+1} -q_{j} )^{3} +\frac{\beta }{4} (q_{j+1} -q_{j} )^{4}   \end{equation} 

where $q_j$ and $p_j$ are the coordinates and conjugate moments, respectively, and N is the number of particles. 

The transformation to normal coordinates is a linear canonical transformation

\[q_{j} =\sum _{k=0}^{N-1}s_{j,k} \xi _{k}^{}  \] 

with

\begin{equation}
 \label{eq19} s_{j,k} =\left\{\begin{array}{l} {\frac{1}{\sqrt{N} } ,\quad \quad \quad \quad \quad \quad k=0} \\ {\sqrt{\frac{2}{N} } \sin (\frac{2\pi kj}{N} +\gamma ),\quad k=1,...[\frac{N-1}{2} ]} \\ {\frac{(-1)^{j} }{\sqrt{N} } ,\quad \quad \quad \quad \quad \quad k=\frac{N}{2} } \\ {\sqrt{\frac{2}{N} } \cos (\frac{2\pi kj}{N} -\gamma ),\quad k=\frac{N}{2} +1,...,N-1} \end{array}\right.
  \end{equation} 

that allows us to write the quadratic part of Hamilton function as the sum of energies of independent oscillators:

\begin{equation} \label{eq20}
 H_{2} =\sum _{k=1}^{N-1}\frac{1}{2} (\eta _{k} ^{2} +\omega _{k}^{2} \xi _{k}^{2} )  
\end{equation} 

where $\xi_k$ and $\eta_k$ are the amplitudes and momentums of the NNMs. (The coordinate corresponding to the center of mass motion ($\xi_0$) is removed from (\ref{eq20})).

\begin{figure}[htbp]
            \centering
        \includegraphics[width=80.7mm, height=56.2mm]{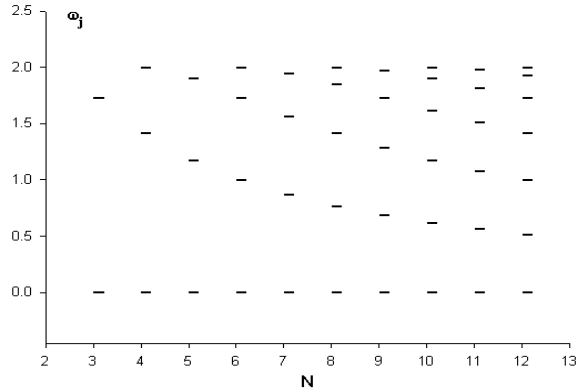}
        \caption{Spectra of eigenvalues of FPU chain with periodic boundary conditions for various number of particles}
	\label{f6}
    \end{figure}

The eigenvalues  

\begin{equation} \label{eq21} 
\omega _{k} =2\sin (\frac{\pi k}{N} ),\quad k=0,\ldots ,N-1 
\end{equation} 

are restricted by the top-frequency $\omega_{N/2}=2$. The zero eigenvalue $\omega_0=0$ corresponds to the motion of mass center. The spectra of periodic FPU chains with various numbers of particles are shown in Fig.\ref{f6}. The short horizontal lines correspond to the eigen-frequencies of the chain with N particles, and all of the eigenvalues in the interval $0<\omega_k<2$ are twice degenerate. Let us note that in the case of the system with even number of particles the gap between frequencies of the top non-degenerate mode k=N/2 and nearby modes decreases fast enough when the number of particles increases. This fact is important for further analysis of instability of top-frequency $\pi $-mode.

One should note that in spite of harmonic part of Hamilton function does not sensitive to the value of the phase $\gamma $ in the canonical transformation (\ref{eq19}), it is not valid for the nonlinear interaction. The nonlinear part of the Hamilton function for the periodic $\beta $-FPU chain has the simplest representation with the specific choice of $\gamma=\pi/4$  ~[\cite{book9}]:

\begin{equation} \label{eq22} 
H_{4} =\frac{\beta }{8N} \sum _{k,l,m,n=1}^{N-1}\omega _{k} \omega _{l} \omega _{m} \omega _{n} C_{k,l,m,n} \xi _{k} \xi _{l} \xi _{m} \xi _{n}  ,  
\end{equation} 

where 

\[C_{k,l,m,n} =-\Delta _{k+l+m+n} +\Delta _{k+l-m-n} +\Delta _{k-l+m-n} +\Delta _{k-l-m+n} \] 

with

\[\Delta _{r} =\left\{\begin{array}{l} {(-1)^{r} ,\; if\quad r=mN,\; m\in  Z } \\ {0\quad \quad otherwise.} \end{array}\right. . \] 

The respective equations of motion are written in the form:

\begin{equation} \label{eq23} 
\frac{d^{2} \xi _{k} }{dt^{2} } +\omega _{k} ^{2} \xi _{k} +\frac{\beta }{2N} \omega _{k} F_{k} (\{ \xi \} )=0,  
\end{equation} 

where

\[F_{k} (\{ \xi \} )=\sum _{l,m,n=1}^{N-1}\omega _{l} \omega _{m} \omega _{n} C_{k,l,m,n} \xi _{l} \xi _{m} \xi _{n}  . \] 

If we assume that parameter of non-linearity $\beta $ is equal to zero, the equations (\ref{eq23}) describe the set of independent oscillators. What happens when the parameter $\beta $ is not equal to zero? Whether some NNMs preserve their independence? It is known ~[\cite{book9}] that the modes with indexes

k=0, N/4, N/3, N/2, 2N/3, 3N/4,

are decoupled from other modes and they are one-mode particular solutions of eqs.(\ref{eq23}). The mode with zero index is trivial. The top-frequency mode with $k=N/2$, which is frequently named as ``$\pi $-mode'' or ``zone-boundary mode'', is one of the exact solution of equations (\ref{eq23}). The remaining modes, which are equidistant from the N/2, form a series of two-mode invariant manifolds in the Fourier space of the system. It is necessary to note that this result is valid for the $\beta $-FPU lattice.

But what we can say about the other types of elementary excitations? It is clear that the linear properties of normal modes have to be approximately preserved if the parameter of nonlinearity $\beta $ is small (or, that is the same, the amplitudes of modes are small enough). The traditional conception is that the linear oscillators far from resonant conditions do not interact. Moreover, the presence of 1:1 resonant conditions (between modes with indexes k and (N-k)) in the periodic weakly nonlinear FPU-chain does not disrupt the integrability of original system ~[\cite{book10},\cite{book11}].

To analyze a dynamics of the system we introduce the complex amplitude of normal mode $\Psi $ as follows:

\[\Psi _{k} =\frac{1}{\sqrt{2} } (\eta _{k} +i\omega _{k} \xi _{k} ),\quad \Psi _{k}^{*} =\frac{1}{\sqrt{2} } (\eta _{k} -i\omega _{k} \xi _{k} ), \] 

in the terms of which the equations (\ref{eq23}) one can be written as the coupled nonlinear equations:

\begin{equation} \label{eq24} 
i\frac{d\Psi _{k} }{dt} +\omega _{k} \Psi _{k} -  \frac{\beta }{8N} \omega _{k} \sum _{l,m,n=1}^{N-1}C_{k,l,m,n} (\Psi _{l} -\Psi _{l}^{*} )(\Psi _{m} -\Psi _{m}^{*} )(\Psi _{n} -\Psi _{n}^{*} )   =0 
\end{equation} 

We need to extract the mono-frequency motion to concentrate our attention on nonlinear evolution. Because we consider the processes in the time scale that is essentially large than $1/\omega_k$, we have to introduce the time hierarchy:

\begin{equation} \label{eq25} 
\tau _{0} =t,\quad \tau _{1} =\varepsilon \tau _{0} ,\quad \tau _{2} =\varepsilon ^{2} \tau _{0} ,  
\end{equation} 

where the ``fast'' time $\tau_0$ corresponds to the natural time scale of the modes and the ``slow'' times $\tau_1$ and $\tau_2$ allow to study the behavior of envelope functions. Keeping in mind the dependence of chain properties from the chain length, one should consider the value $1/N$ as a small parameter $\varepsilon $.

Let us introduce the envelope function $\varphi_k$ as follows:

\[\Psi _{k} =\varphi _{k} e^{i\omega _{k} t} . \]

\begin{equation} \label{eq26} 
\varphi _{k} =\sqrt{\varepsilon } (\chi _{k,1} +\varepsilon \chi _{k,2} +\varepsilon ^{2} \chi _{k,3} +...).  
\end{equation} 

The multi-scale expansion based on the equations (\ref{eq25}, \ref{eq26}) allows us to construct the dynamical equations for the main-order approximation amplitudes $\chi_{k,1}$. One can show that for the $\beta $-FPU system equations (\ref{eq24}) admit the particular solutions in the form of a pair of degenerate modes with indexes (k, N-k) when the eigenvalues $\omega_k$ are different enough. We call these degenerate modes as ``conjugate'' ones.

\subsection*{Interaction of conjugate modes.}

As it was mentioned above, in the small FPU chain the modes, which are not conjugate, have the frequencies which are differed significantly. In such a case we consider first the interaction between conjugate modes. One can see that the amplitudes of main-order approximation $\chi_{k,1}$ are independent on the ``fast time'' $\tau_0$. Really, in the lower-order by small parameter $\epsilon$ we get:

\[
\epsilon^{1/2}: \begin{array}{l} {i\frac{\partial \chi _{k,1} }{\partial \tau _{0} } =0}\\
 {i\frac{\partial \chi _{N-k,1} }{\partial \tau _{0} } =0} \end{array}.
\] 

The equations of the next order by small parameter $\epsilon$ are read as follows:

\begin{equation} \label{eq27}
\epsilon^{3/2}: \begin{array}{l} {i\frac{\partial \chi _{k,2} }{\partial \tau _{0} } +i\frac{\partial \chi _{k,1} }{\partial \tau _{1} } =0} \\
 {i\frac{\partial \chi _{N-k,2} }{\partial \tau _{0} } +i\frac{\partial \chi _{N-k,1} }{\partial \tau _{1} } =0} \end{array}. 
\end{equation}

The appropriate solutions of equations (\ref{eq27}) are

\[\begin{array}{l} {\chi _{k,2} =0,\quad \frac{\partial \chi _{k,1} }{\partial \tau _{2} } =0} \\ {\chi _{N-k,2} =0,\quad \frac{\partial \chi _{N-k,1} }{\partial \tau _{2} } =0} \end{array}. \] 

So the equations of the next order by small parameter are:

\begin{equation} \label{eq28}
\epsilon^{5/2}: \begin{array}{l} {i(\frac{\partial \chi _{k,1} }{\partial \tau _{2} } +\frac{\partial \chi _{k,3} }{\partial \tau _{0} } )+(\kappa _{1} |\chi _{k,1} |^{2} +\kappa _{2} |\chi _{N-k,1} |^{2} )\chi _{k,1} +} \\ {\quad \quad \quad \quad \quad \quad \kappa _{3} \chi _{N-k}^{2} \chi _{k}^{*} =R_{k} } \\ {i(\frac{\partial \chi _{N-k,1} }{\partial \tau _{2} } +\frac{\partial \chi _{N-k,3} }{\partial \tau _{0} } )+(\kappa _{1} |\chi _{N-k,1} |^{2} +\kappa _{2} |\chi _{k,1} |^{2} )\chi _{N-k,1} +} \\ {\quad \quad \quad \quad \quad \quad \kappa _{3} \chi _{k,1}^{2} \chi _{N-k,1}^{*} =R_{N-k} } \end{array}. 
\end{equation}

The right hand side items $R_j$ in equations (\ref{eq28}) contain both the ``resonant'' and the ``non-resonant'' parts of the interactions that dependent on the fast time $\tau_0$. The spectrum of FPU lattice with periodic boundary conditions contains only two types of resonances. The first one is the 1:1 resonance between conjugate modes (k, N-k) and the second one is 2:1 resonance between $\pi $-mode and the modes with indexes N/3 and 2N/3. Therefore, the right hand sides of equations \ref{eq28} contain the terms similar to $\chi _{m} \chi _{n}^{*} \chi _{j} $ with indexes ${(m,\; n)}\ne {(k,\; N-k)}$ and j=k or N-k. After integrating these equations over the fast time the requirement of the absence of any secular terms leads to that only the terms mentioned remain in the right hand sides. If we take an interest in a mutual dynamics of conjugate modes, one should choose the initial conditions for equations (\ref{eq28}) as follows:

\[\chi _{j,1} (\tau =0)=0,\quad j\ne k,\; N-k. \] 

In such a case the right hand part items in equations (\ref{eq28}) are equal to zero and the resulting equations are uncoupled with any other modes:

\begin{equation} \label{eq29} 
\begin{array}{l} {i\frac{\partial \chi _{k,1} }{\partial \tau _{2} } +(\kappa _{1} |\chi _{k,1} |^{2} +\kappa _{2} |\chi _{N-k,1} |^{2} )\chi _{k,1} +\kappa _{3} \chi _{N-k,1}^{2} \chi _{k,1}^{*} =0} \\ {i\frac{\partial \chi _{N-k,1} }{\partial \tau _{2} } +(\kappa _{1} |\chi _{N-k,1} |^{2} +\kappa _{2} |\chi _{k,1} |^{2} )\chi _{N-k,1} +\kappa _{3} \chi _{k,1}^{2} \chi _{N-k,1}^{*} =0} \end{array}.  
\end{equation}

\[\kappa _{1} =\frac{3\beta \omega _{k} }{8} C_{k,k,k,k} ,\quad \kappa _{2} =\frac{3\beta \omega _{k} }{4} C_{k,k,N-k,N-k} ,\quad \kappa _{3} =\frac{3\beta \omega _{k} }{8} C_{k,k,N-k,N-k} . \] 

One can easily  check that the coefficients $C_{k,k,k,k}$ and $C_{k,k,N-k,N-k}$ have the following values:

\[C_{k,k,k,k} =\left\{\begin{array}{l} {3,\quad k\ne N/4} \\ {4,\quad k=N/4} \end{array}\right. ;\quad C_{k,k,N-k,N-k} =\left\{\begin{array}{l} {1,\quad k\ne N/4} \\ {0,\quad k=N/4} \end{array}\right. . \] 

Such values of parameters show that the equations have the universal structure for any pair of conjugate modes. The only exceptions are the modes with indexes (N/4, 3N/4), which form the one-solution manifolds in the exact problem [~\cite{book9}]. Moreover, the following relation between coefficients in equations (\ref{eq29}) is valid:

\begin{equation} \label{eq30} 
\kappa _{1} -\kappa _{2} -\kappa _{3} =0,  
\end{equation} 

which is significant as we will see later. One can assert that this relation is valid for $\alpha \beta $-FPU lattices too. 

The Hamilton function corresponding to equations (\ref{eq29}):

\begin{equation} \label{eq31} 
\begin{array}{l} {H_{\chi } =\frac{\kappa _{1} }{2} (|\chi _{k} |^{4} +|\chi _{N-k} |^{4} )+\kappa _{2} |\chi _{k} |^{2} |\chi _{N-k} |^{2} +} \\ {\quad \quad \frac{\kappa _{3} }{2} (\chi _{k}^{* 2} \chi _{N-k}^{2} +\chi _{k}^{2} \chi _{N-k}^{* 2} )} \end{array}.  
\end{equation} 

It is easily to show that equations (\ref{eq29}) besides the integrals energy and momentum have an additional integral of motion

\begin{equation} \label{eq32} 
X_{k} =|\chi _{k} |^{2} +|\chi _{N-k} |^{2} .  
\end{equation} 

Here we discard the second term in the index of modes. The value $X_k$ is analogous to the quantum representation and it is often named as ``occupation number''. It characterizes a full level of excitation of the conjugate modes. Taking into account the relation (\ref{eq32}) one can rewrite the equations (\ref{eq29}) as follows:

\begin{equation} \label{eq33} 
\begin{array}{l} {i\frac{\partial \chi _{k} }{\partial \tau _{2} } +\kappa _{1} X_{k} \chi _{k} +\kappa _{3} (\chi _{N-k} \chi _{k}^{*} -\chi _{k} \chi _{N-k}^{*} )\chi _{N-k} =0} \\ {i\frac{\partial \chi _{N-k} }{\partial \tau _{2} } +\kappa _{1} X_{k} \chi _{N-k} +\kappa _{3} (\chi _{k} \chi _{N-k}^{*} -\chi _{N-k} \chi _{k}^{*} )\chi _{k} =0} \end{array}.  
\end{equation} 

The first items in the equations (\ref{eq33}) describe the change of eigenvalue $\omega_k$ caused by excitation of the conjugate modes. The second terms describe the ``exchange'' interaction between modes. It is clear that equations (\ref{eq33}) admit the one-mode solutions:

\[
\left\{\begin{array}{l} {\chi _{k} =\sqrt{X_{k} } e^{i\Omega \tau_{2} } } \\ {\chi _{N-k} =0} \\ {X_{k} =|\chi _{k} |^{2} ,\Omega =\kappa _{1} X_{k} } \end{array}\right. \quad \quad \left\{\begin{array}{l} {\chi _{k} =0} \\ {\chi _{N-k} =\sqrt{X_{k} } e^{i\Omega \tau_{2} } } \\ {X_{k} =|\chi _{N-k} |^{2} ,\Omega =\kappa _{1} X_{k} } \end{array}\right. . 
\] 

If the initial conditions include the excitation of both modes, one can show that the value $iG_{k} ^{-} =(\chi _{N-k} \chi _{k}^{*} -\chi _{N-k}^{*} \chi _{k} )$ is the integral of motion too. The energy of the system can be written as the function of $X_k$ and $G_{k}^{-}$:

\[E_{k} =\frac{\kappa _{1} }{2} X_{k}^{2} -\frac{\kappa _{3} }{2} G_{k}^{- 2} . \] 

Therefore the equations (\ref{eq33}) can be  linearized:

\begin{equation} \label{eq34} 
\begin{array}{l} {i\frac{\partial \chi _{k} }{\partial \tau _{2} } +\kappa _{1} X_{k} \chi _{k} -i\kappa _{3} G_{k} ^{-} \chi _{N-k} =0} \\ {i\frac{\partial \chi _{N-k} }{\partial \tau _{2} } +\kappa _{1} X_{k} \chi _{N-k} +i\kappa _{3} G_{k} ^{-} \chi _{k} =0} \end{array}.  
\end{equation} 

The particular solution of equations (\ref{eq34}) is

\begin{equation} \label{eq35} 
\begin{array}{l} {\chi _{k} =a_{k} \exp [i(\Omega \tau _{2} +\frac{\Delta }{2} )]} \\ {\chi _{N-k} =a_{N-k} \exp [i(\Omega \tau _{2} -\frac{\Delta }{2} )]} \end{array} 
\end{equation} 

with

\[\Omega =\kappa _{1} X_{k} +\kappa _{3} G_{k} ^{-} . \] 

and

\[a_{k} =\pm a_{N-k} =\sqrt{\frac{X_{k} }{2} } ,\quad \Delta =\pm {\raise0.7ex\hbox{$ \pi  $}\!\mathord{\left/{\vphantom{\pi  2}}\right.\kern-\nulldelimiterspace}\!\lower0.7ex\hbox{$ 2 $}} . \] 

This solution corresponds to the elliptical in-phase or anti-phase mode. The general solution can be presented in the form:

\begin{equation} \label{eq36} 
\begin{array}{l} {\chi _{k} =\sqrt{X_{k} } e^{i\kappa _{1} X_{k} \tau _{2} } \cos \theta e^{i\Delta /2} } \\ {\chi _{N-k} =\sqrt{X_{k} } e^{i\kappa _{1} X_{k} \tau _{2} } \sin \theta e^{-i\Delta /2} } \end{array}.  
\end{equation} 

The equations (\ref{eq34}) in the terms of variables $\theta$ and $\Delta $ have the following form:

\begin{equation} \label{eq37} 
\begin{array}{l} {\sin 2\theta [\frac{d\theta }{d\tau _{2} } +\kappa _{3} G_{k} ^{-} \cos \Delta ]=0} \\ {\sin 2\theta [\frac{d\Delta }{d\tau _{2} } -2\kappa _{3} G_{k} ^{-} ctg2\theta \sin \Delta ]=0} \end{array}.  
\end{equation} 

At that, the parameter $G_{k}^{-}$ is connected with the variables ($\Delta $, $\theta$) by relation:

\[\sin 2\theta \sin \Delta =G_{k} ^{-} /X_{k} =const. \] 

The behavior of the system can be analyzed efficiently in the reduced phase plane ($\Delta $, $\theta$) (see Fig. \ref{f2}).

\begin{figure}[htbp]
            \centering
        \includegraphics[width=84.4mm, height=61.8mm]{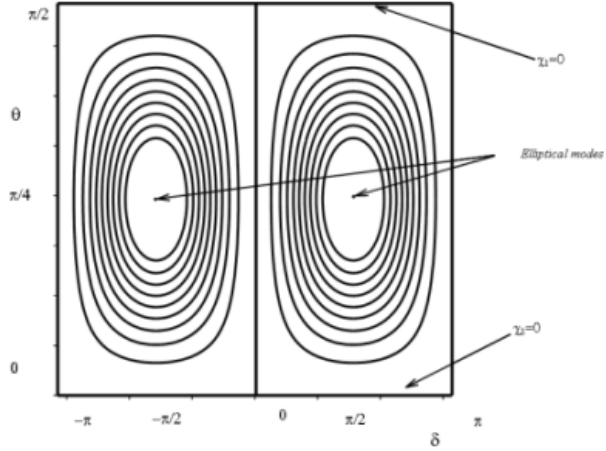}
        \caption{Phase plane of equations (\ref{eq37}) in the terms ($\Delta $, $\theta$). The stationary point ($\Delta =\pm\pi/2$, $\theta=\pi/4$) corresponds to elliptical mode ($G_k^{-}=X_k$). The rectangular trajectories are the LPTs and they correspond to $G_k^{-}=0$. Both presented quadrants are equivalent ones because of periodicity conditions.}
	\label{f7}
    \end{figure}

There are two characteristic phase trajectories in Fig.\ref{f7}. First of them is the central point ($G_{k}^{-}/X_k=1$, $\Delta =\pi/2$, $\theta=\pi/4$) corresponding to elliptical normal modes (\ref{eq35}). The trajectories close to this point correspond to weak intermode energy exchange. Second trajectory ($G_k^-/X_k=0$) is limiting phase trajectory (LPT) describing complete energy transfer. Let us note that the rate of energy exchange rises with increasing G1- and is minimal if $G_k^-/X_k=0$. So, both intensity and rate of energy transfer are controlled by the parameter $G_{k}^{-}/X_k$. The most intensive energy exchange is attained for the trajectories close to LPT (for LPT itself the rate is equal to zero that is strongly different from the case of two linearly coupled nonlinear oscillators ~[\cite{book5}], where complete interparticle energy exchange occurs with finite rate).

\begin{figure}[htbp]
            \centering
        (a)\includegraphics[width=60.mm, height=47.7mm]{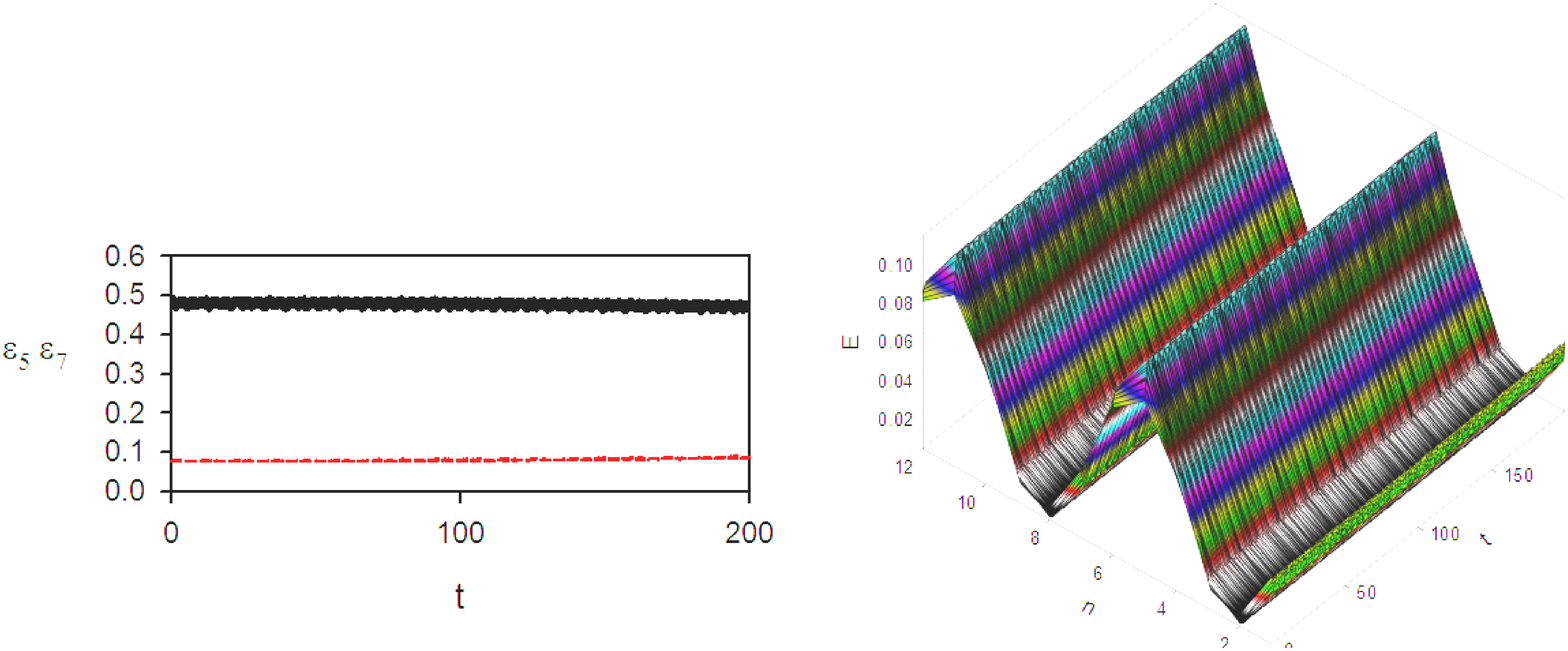}
	(b)\includegraphics[width=60.mm, height=47.7mm]{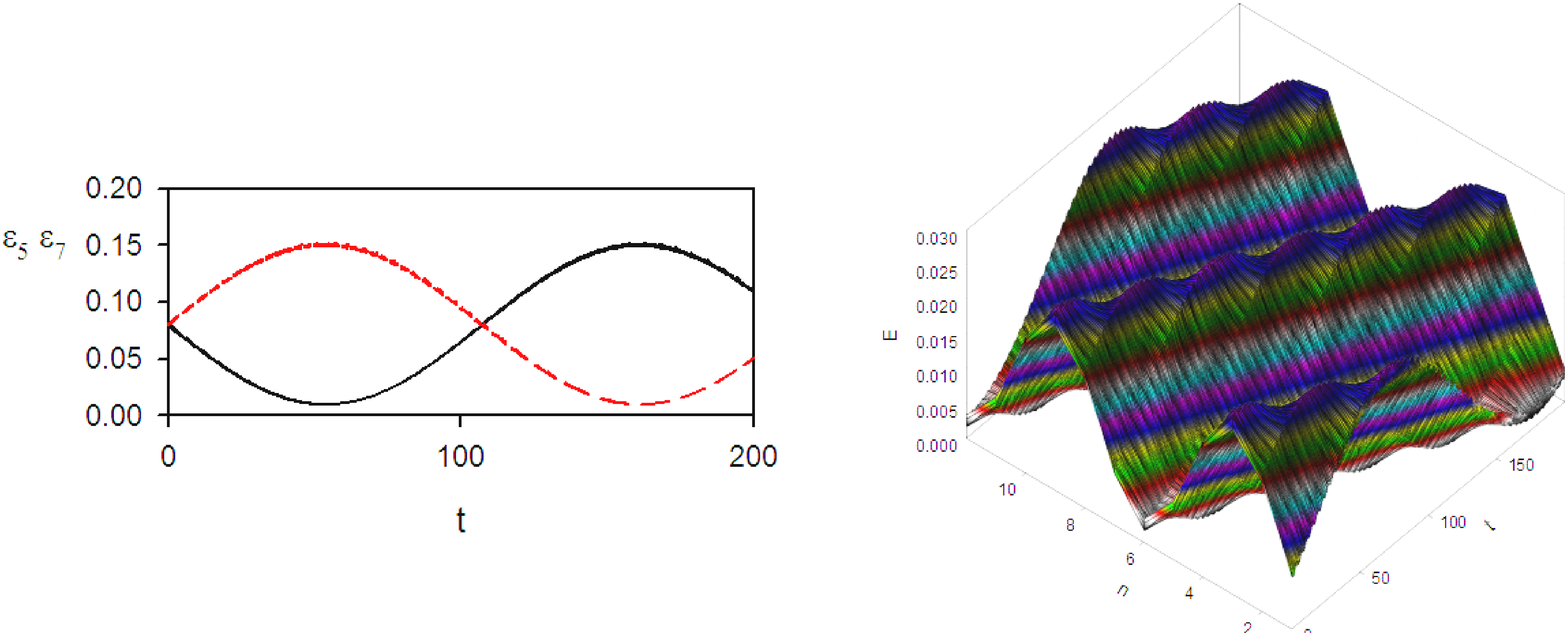}
        \caption{Effect of energy exchange between conjugate modes (5,7) in the $\beta $-FPU with 12 particles. (a) The initial excitation corresponds to $G_k^-/X_k=0$. The absence of energy exchange between modes (left graph) leads to immobility of energy along the chain (right graph). (b) The initial excitation corresponds to $G_k^-/X_k\sim 0.1$. The intensive energy exchange between high-frequency conjugate modes (left graph) leads to excitation mobility through the chain (right graph).}
	\label{f8}
    \end{figure}

Certain results of computer simulation of high frequency part of normal modes in the $\beta$--FPU chain with 6 particles are shown in figure \ref{f8}.

Up to this time we have considered the dynamics of conjugate modes with initial phase shift $\gamma =\pi /4$ (see eq.(\ref{eq19})). One can easy to see that the harmonic part of Hamilton function (\ref{eq18}) does not depend on the choice of $\gamma $ because any superposition of linear normal modes is a normal mode too. In the case of NNMs this statement is not more valid. Therefore, it is very interesting to estimate the effect of change of phase shift on the dynamics of NNMs. Let us consider the conjugate modes with indexes (N/4, 3N/4), that form the one-mode solutions in the exact nonlinear problem ~[\cite{book9}]. Therefore, the equations of nonlinear dynamics (\ref{eq29}) do not contain any exchange terms, if the phase shift is equal to $\pi/4$:

\[
\begin{array}{l} {i\frac{\partial \chi _{N/4} }{\partial \tau _{2} } +\kappa _{1} |\chi _{N/4} |^{2} \chi _{N/4} =0} \\ {i\frac{\partial \chi _{3N/4} }{\partial \tau _{2} } +\kappa _{1} |\chi _{3N/4} |^{2} \chi _{3N/4} =0} \end{array}. 
\] 

For the $\beta $-FPU lattices the phase shift $\gamma =0$ which does not coincide with $\pi /4$, leads to the equations those are like the equations (\ref{eq34}). If $\gamma=0$ is our choice for phase shift it is turns out that the relation 

\begin{equation} \label{eq38} 
\kappa _{1} -\kappa _{2} +\kappa _{3} =0 
\end{equation} 

is valid instead of eq.(\ref{eq35}). It leads to that the value 

\[G_{N/4} ^{+} =(\chi _{n/4}^{*} \chi _{3N/4} +\chi _{N/4} \chi _{3N/4}^{*} )\] 

is the integral of motion instead of the parameter $G_{k}^{-}$. In such a case the linearized form of equations (\ref{eq33} )is:

\begin{equation} \label{eq39} 
\begin{array}{l} {i\frac{d\chi _{N/4} }{d\tau _{2} } +\kappa _{1} X_{N/4} \chi _{N/4} +\kappa _{3} G_{N/4} ^{+} \chi _{3N/4} =0} \\ {i\frac{d\chi _{3N/4} }{d\tau _{2} } +\kappa _{1} X_{N/4} \chi _{3N/4} +\kappa _{3} G_{N/4} ^{+} \chi _{3N/4} =0} \end{array}.  
\end{equation} 

The stationary solutions have the following form:

\begin{equation} \label{eq40} 
\begin{array}{l} {\chi _{N/4} =\sqrt{\frac{X_{N/4} }{2} } \exp [i(\Omega \tau _{2} +\frac{\Delta _{} }{2} )]} \\ {\chi _{3N/4} =\sqrt{\frac{X_{N/4} }{2} } \exp [i(\Omega \tau _{2} -\frac{\Delta }{2} )]} \\ {\Omega =\kappa _{1} X_{N/4} +\kappa _{3} G_{N/4} ^{+} } \end{array} 
\end{equation} 

with

\[\Delta =0,\quad G_{N/4} ^{+} =X_{N/4} \] 

or

\[\Delta =\pi ,\quad G_{N/4} ^{+} =-X_{N/4} . \] 

\begin{figure}[htbp]
		\centering
		(a) \includegraphics[width=60.0mm, height=65.0mm] {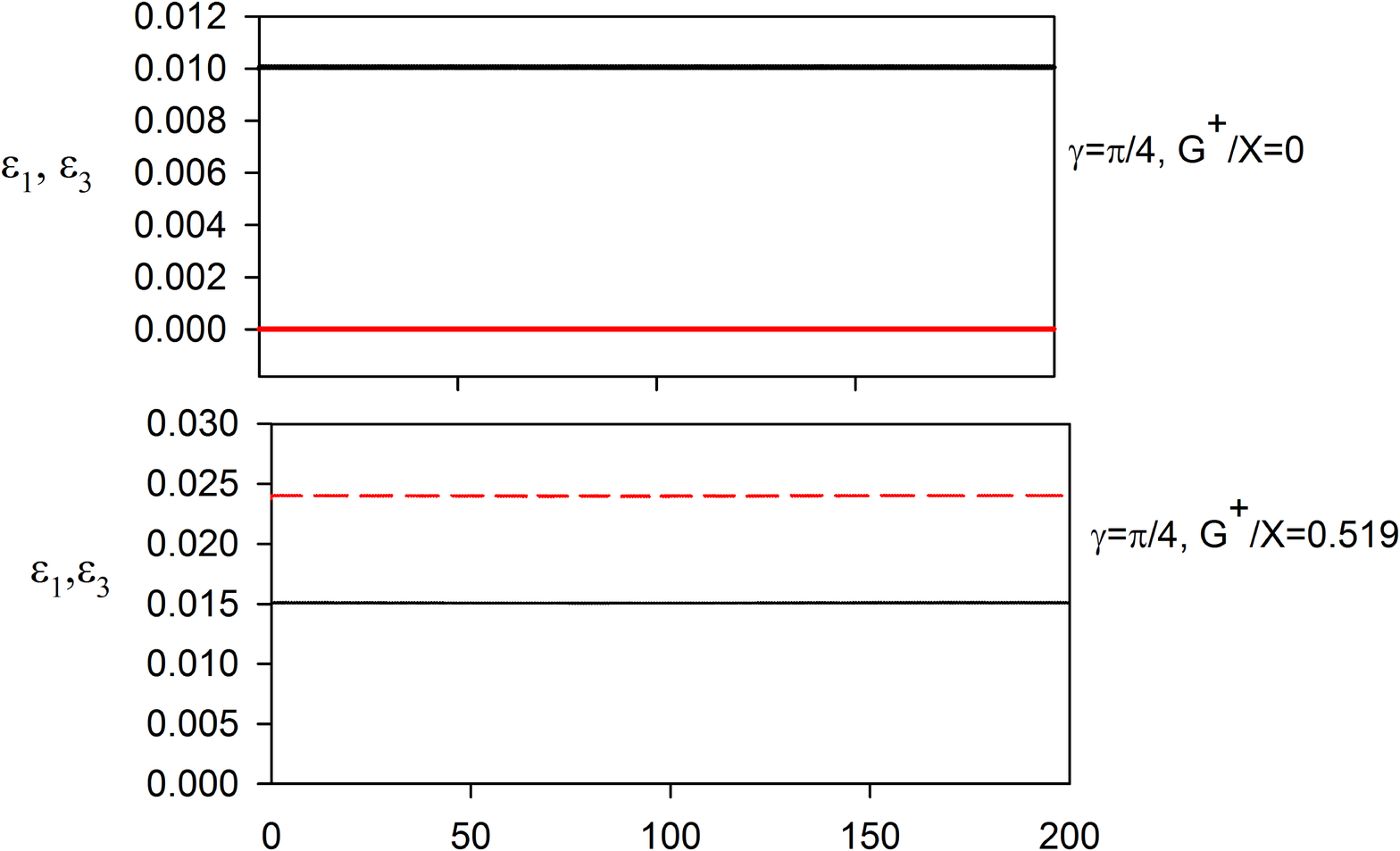}
		(b) \includegraphics[width=60.0mm, height=65.0mm] {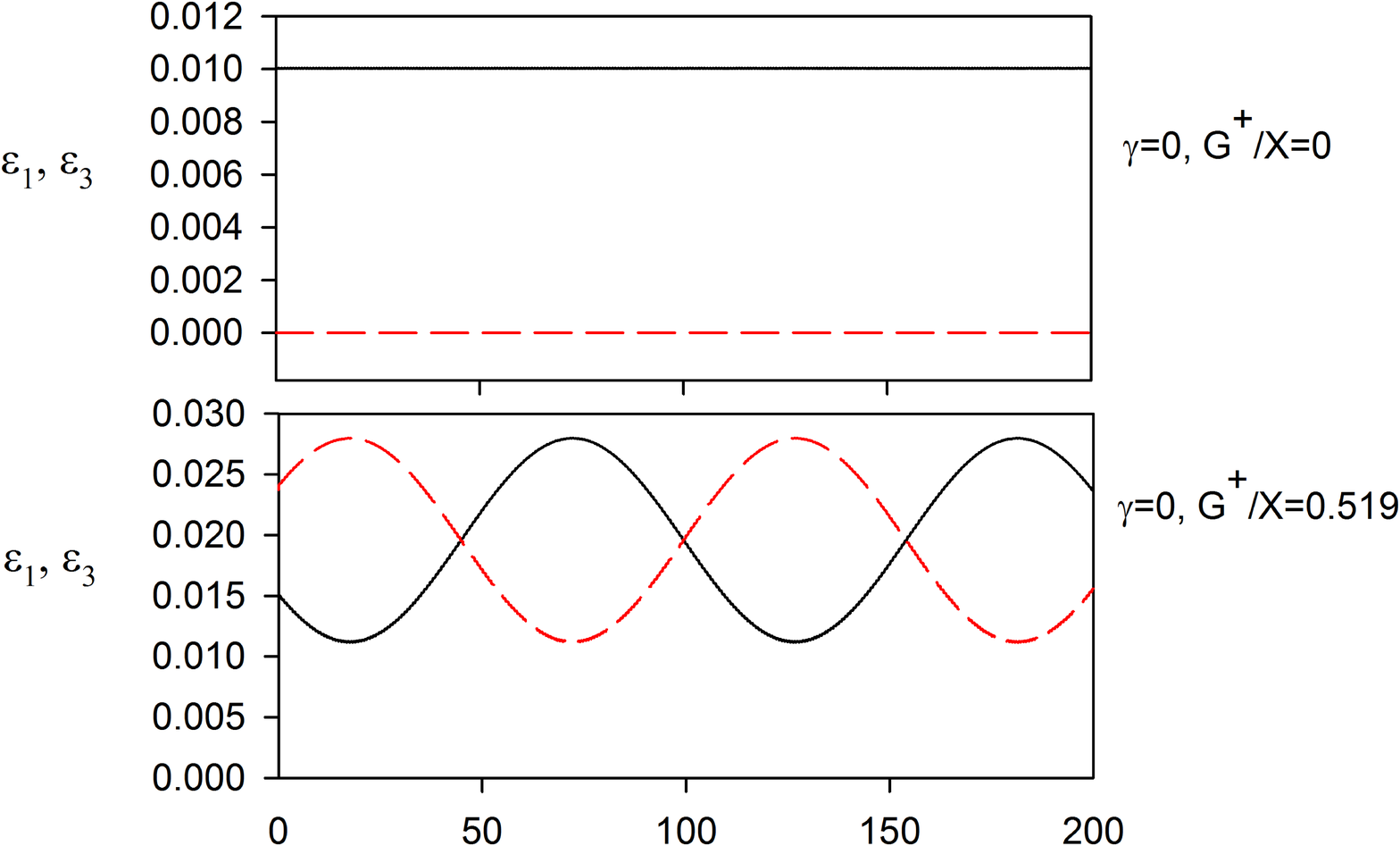}
		\caption{The comparison of energy exchange between normal modes with different phase shift $ \gamma $ in the $ \beta $-FPU chains with 4 particles. (a) The energies of one- and two-mode solutions at $ \gamma = \pi /4 $ . There is no inter-mode energy exchange. (b) If $ \gamma = 0 $ the intensive energy exchange occurs at the same initial conditions. }
		\label{f9}
	\end{figure}

The main difference of these solutions from (\ref{eq35}) is that the displacement of stationary points from the positions ($\pi /4, ~\pi  /2$) and ($\pi /4, ~-\pi/2$) into ($\pi/4,~0$) and ($\pi/4,~\pi $) in the reduced phase plane ($\Delta ,~\theta $). The presence of exchange terms in the equations leads to that the trajectories, starting near the LPT describe the full energy exchange between modes. The comparison of dynamics of normal modes with different $\gamma $ is shown in the fig.\ref{f9}.

Equations (\ref{eq39}) in the terms of variables ($ \Delta , ~\theta $) are written as

\begin{equation} \label{eq41} 
\begin{array}{l} {\sin 2\theta [\frac{d\theta }{d\tau _{2} } +\kappa _{3} G_{N/4} ^{+} \sin \Delta ]=0} \\ {\sin 2\theta [\frac{d\Delta }{d\tau _{2} } +2\kappa _{3} G_{N/4} ^{+} ctg2\theta \cos \Delta ]=0} \end{array}  
\end{equation} 

The last equations determine a relation between angle variables in the form:

\[\sin 2\theta \cos \Delta =G_{N/4} ^{+} /X_{N/4} =const. \] 

\begin{figure}[htbp]
            \centering
        \includegraphics[ width=60.0mm, height=54.1mm]{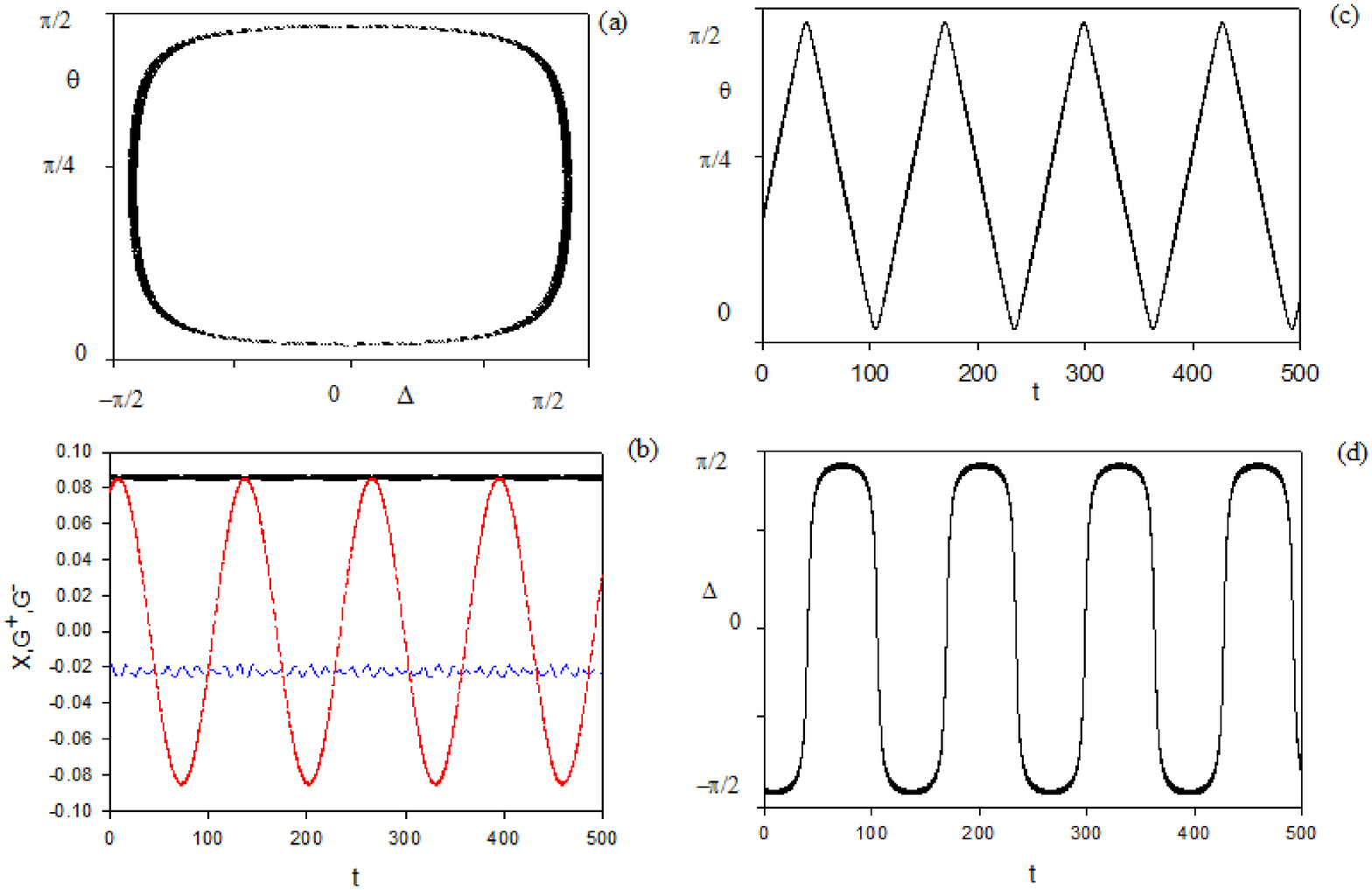}
        \includegraphics[ width=60.0mm, height=54.1mm]{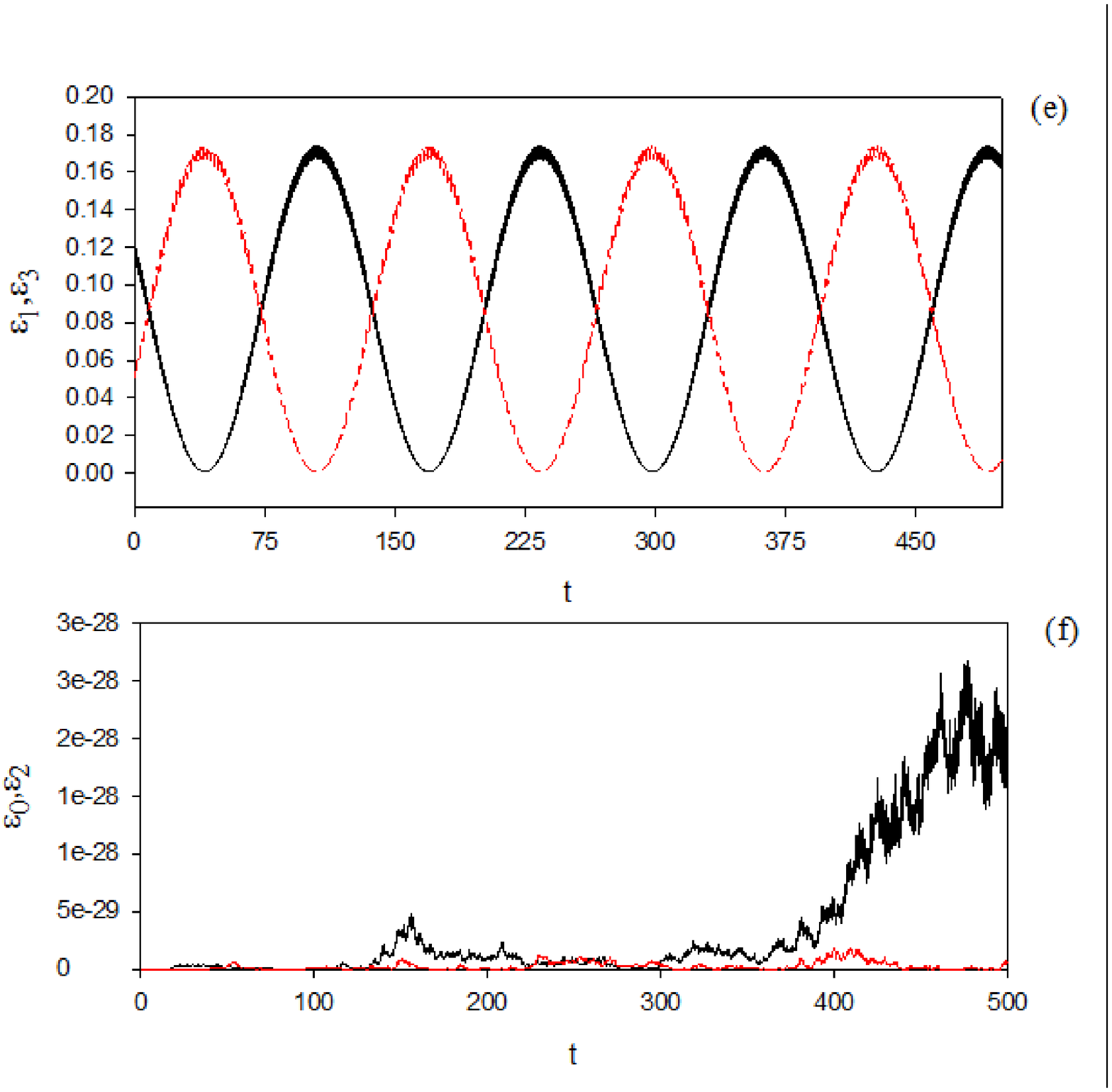}
        \caption{$\beta $-FPU chain with 4 particles: (a) -- ($\Delta , ~\theta $)-plane: the starting point is near the LPT; (b) the values $X$ (black solid), $G^{+}$ (blue dashed) ,$G^{-}$ (red dashed) versus time; (c,d) the time evolution of angles $\theta $ and $\Delta $ (e,f) inter-mode energy exchange.}
	\label{f10}
    \end{figure}

In this case the central point $G_{N/4}^{+}/X_{N/4}$ =1 corresponds to ``supernormal mode'' which is the combination of both resonant modes with identical weights. As for LPT ($G_{N/4}^{+}/X_{N/4} =0$), it describes again a complete intermode transfer with zero rate, so that actual slow energy transfer may be observed for the trajectories close to LPT. The data of computer simulation study confirm the analytical results (see fig. \ref{f10}).

The $\alpha\beta $-FPU chain is more complex object for analytical study. The structure of the phase plane depends on both parameter $\alpha $ and the number of particles in the chain. We use the $\alpha \beta $-FPU chain with 4 particles to clarify the main peculiarities of the system. The parameters of equations (\ref{eq33}) are written as follows:

\[\kappa _{1} =\frac{3\beta }{4} ;\quad \kappa _{2} =\frac{3\beta }{2} ;\quad \kappa _{3} =\frac{3\beta -4\alpha ^{2} }{4} . \] 

One can see that in such lattices the relations (\ref{eq30},\ref{eq38}) are not more valid. It has two consequences: (1) both the parameter $G^{+}$ and the parameter $G^{-}$ are not a constant of motion, and (2) the coefficient $\kappa_3$ can be equal to zero at $\alpha =\sqrt{3\beta /4} $. All of that lead to more complicated description than discussed above. In such a case equations (\ref{eq33}) cannot be linearized, but they are written in the terms of angular variables as before:

\begin{equation} \label{eq42} 
\begin{array}{l} {\sin 2\theta [\frac{d\theta }{d\tau _{2} } +\frac{1}{2} \kappa _{3} X\sin 2\Delta ]=0} \\ {\sin 2\theta \{ \frac{d\Delta }{d\tau _{2} } -X\cos 2\theta [(\kappa _{1} -\kappa _{2} )-\kappa _{3} \cos 2\Delta ]\} =0} \end{array}.  
\end{equation} 

The first integral of equations (\ref{eq42}) defines the relationship between variables ($\Delta $, $\theta$):

\[\sin 2\theta \sqrt{|\frac{\kappa _{1} -\kappa _{2} }{\kappa _{3} } -\cos 2\Delta |} =C=const. \] 

Here the phase plane has a structure depending on the relationship between the constants $\kappa_1$, $\kappa_2$, $\kappa_3$, which are controlled by the parameter $\alpha$ of the potential asymmetry. The case $\alpha$=0 leads to relationship (\ref{eq38}) with supernormal mode as stationary point (fig.\ref{f10}.a). Any trajectory close to LPT corresponds to full energy exchange between modes $\chi_1$ and $\chi_2$ with large period. But occurrence of any asymmetry of potential function leads to separatrix creation.

\begin{figure}[htbp]
            \centering
        \includegraphics[ width=80.mm, height=92mm]{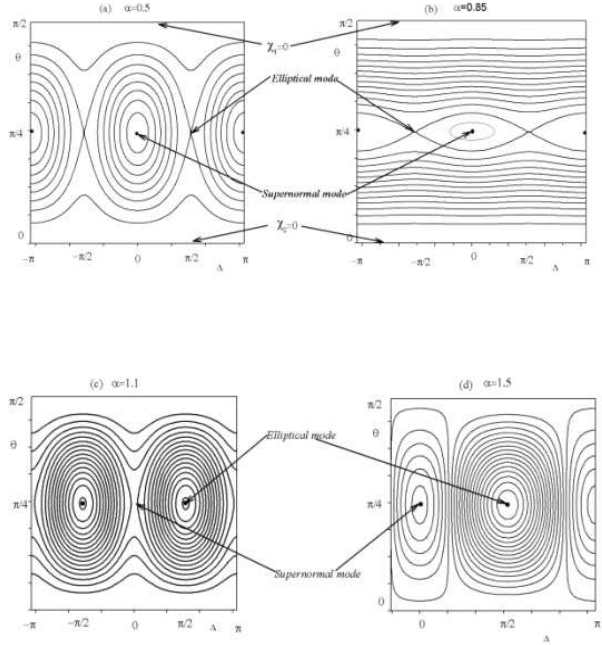}
        \caption{Transformation of the phase plane of equations (\ref{eq42}) with a variation of potential asymmetry parameter $\alpha$ (see text).}
	\label{f11}
    \end{figure}

The singular (saddle) point for small values $0<\kappa_3<\kappa_1-\kappa_2$ (small $\alpha$ values) corresponds to unstable elliptical mode ($\Delta =\pi/2$, $\theta=\pi/4$). As a result we have got a phase plane shown in fig. (\ref{f11}.a). The separatrix crossing the point ($\Delta =\pi/2$, $\theta=\pi/4$) separates the domain of closed trajectories with partial energy exchange from the transit-time trajectories for which an energy exchange is practically absent. The domain of closed trajectories decreases while the parameter $\kappa_3$ tends to zero at $\alpha=0.866$ (fig. \ref{f11}.b). Two branches of separatrix are reduced to straight lines and the phase plane is filled with transit-time trajectories, which are mutually parallel ones. The supernormal mode ($\Delta =0$, $\theta=\pi/4$) appears as the saddle point within the interval --($\kappa_1-\kappa_2)<\kappa_3<0$ for $0.866<\alpha<1.2248$ (fig. \ref{f11}.c). And there are two stationary points ($\Delta  =\pi/2$, $\theta=\pi/4$) and ($\Delta =0$, $\theta=\pi/4$) when the parameter $\kappa_3<(\kappa_1-\kappa_2$) ($\alpha>1.2248$) (fig. \ref{f11}.d). In such a case, the trajectories rounding both supernormal and elliptic stationary points and close to the LPT, correspond to intensive energy exchange between conjugate normal modes. So, the dynamics of the particles drastically changes with a variation of parameter $\kappa_3$, which, in turn, is determinate by the asymmetry of potential.

\begin{figure}[htbp]
            \centering
        \includegraphics[width=81.0mm, height=85.0mm]{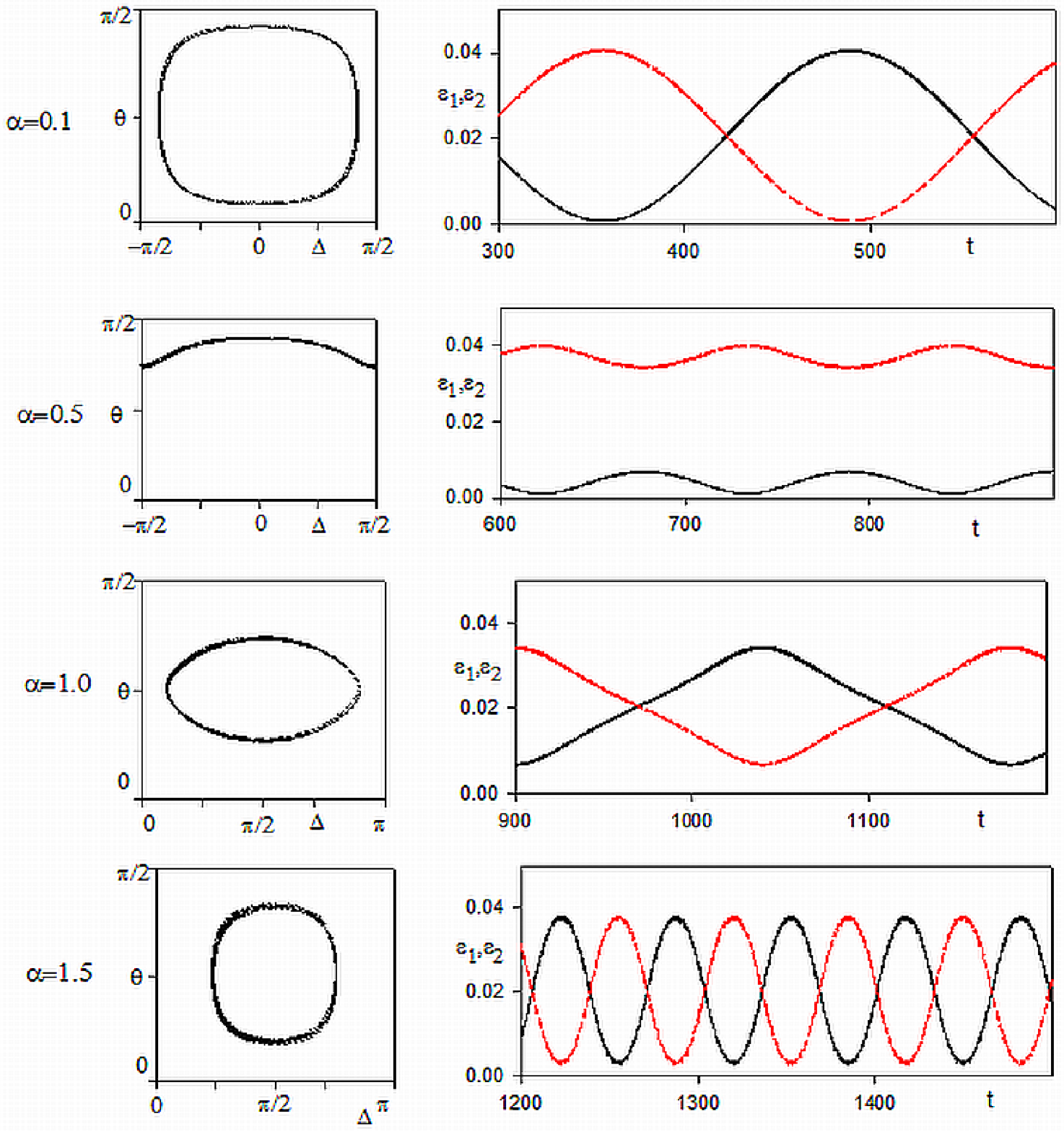}
        \caption{Intermode energy exchange in the $\alpha $$\beta $-FPU chain with 4 particles with various parameter $\alpha $. The left column contains the phase trajectories in the ($\Delta , \theta $)-plane and the right column show the energy exchange between conjugate modes. }
	\label{f12}
    \end{figure}

To check the main analytical results we have performed the computer simulations of dynamics of several FPU-chains with various number of particles (N=3,\dots ,8). Some results of these simulations in the partial case N=4 are shown in the fig. \ref{f12}. The computer simulation was continued since t=0 up to t=1500, but the parameter of asymmetry was changed at t=300, 600, 900, and 1200. The starting points are inside the separatrix at $\alpha =0.1$ and $\alpha =1.0$, but it is in the transit-time area for $\alpha =0.5$. The last means that the energy exchange between modes is forbidden. Because separatrix is absent at $\alpha =1.5$ the complete energy exchange corresponds any trajectories starting near the LPT.

\begin{figure}[htbp]
            \centering
        (a) \includegraphics[width=40.4mm, height=60.0mm]{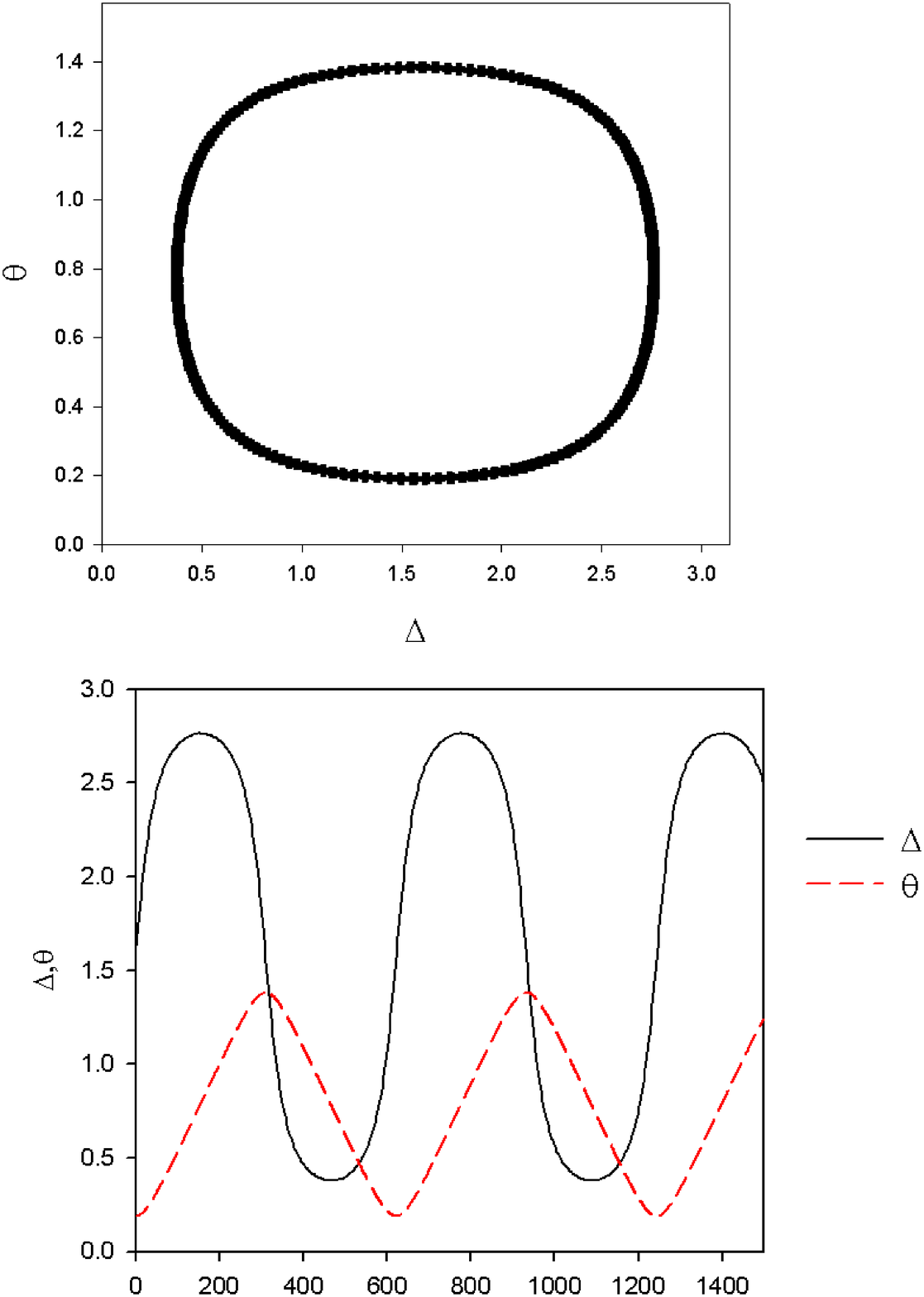} 
        (b) \includegraphics[width=56.6mm, height=60.0mm]{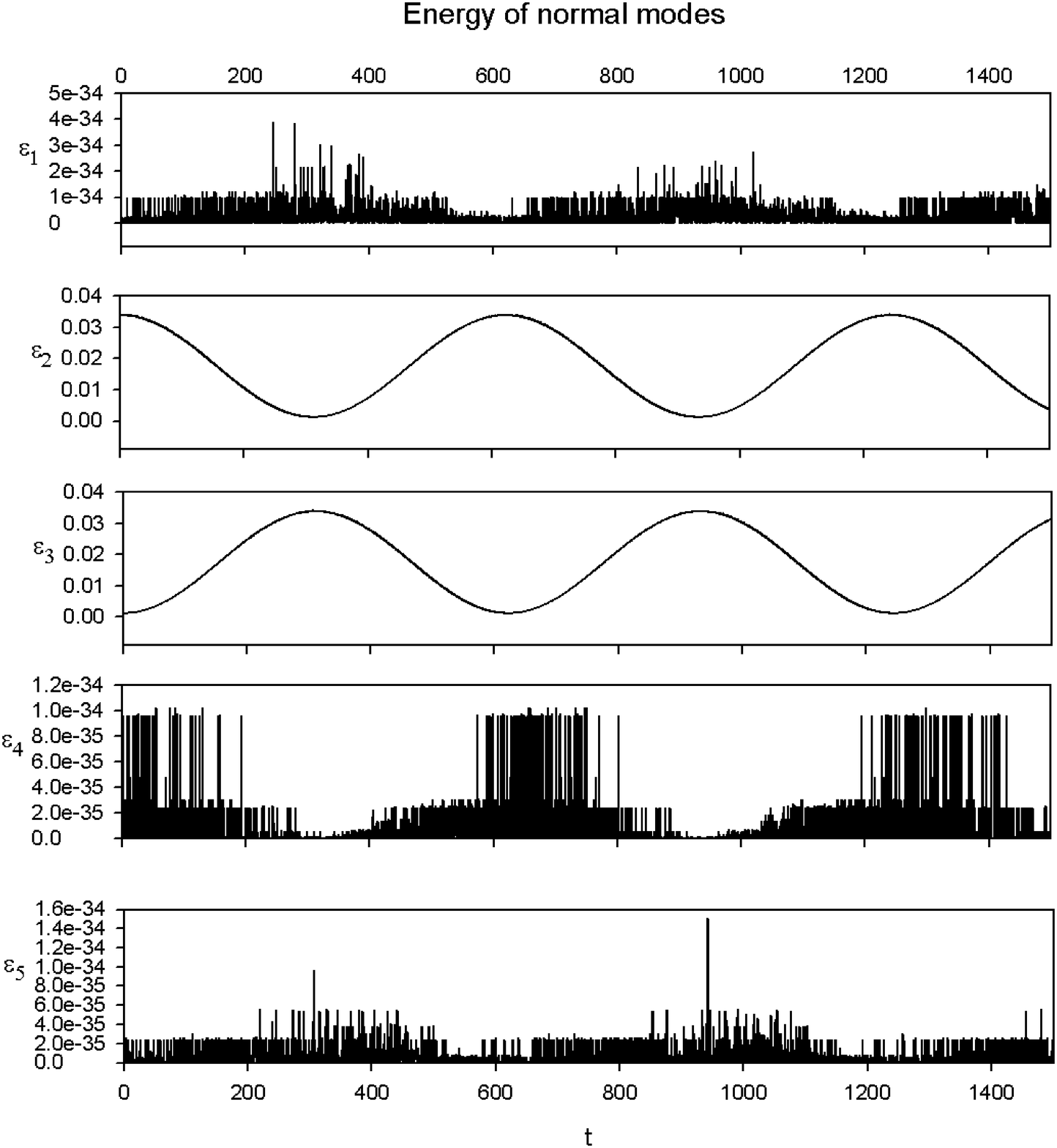}
        \caption{FPU chain with 6 particles: (a) Phase trajectory and angles $\Delta $ and $\theta $; 
	(b) energy of conjugated modes.}
	\label{f13}
\end{figure}

\begin{figure}[htbp]
            \centering
        \includegraphics[width=60.6mm, height=84.4mm]{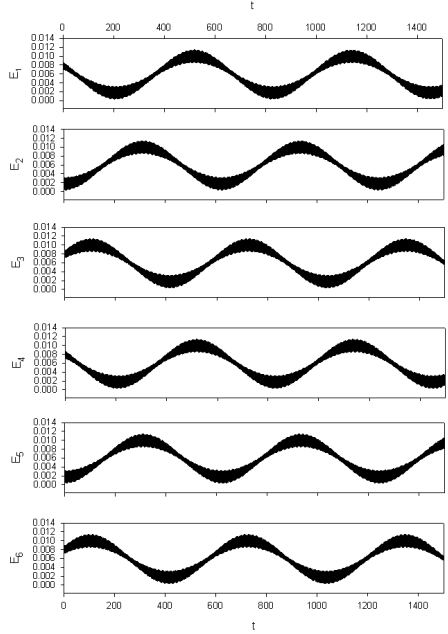}
        \caption{FPU chain with 6 particles: the energy transfer along the chain corresponds to the data shown in the fig.\ref{f13}. $E_n$ -- energy of the particles.}
	\label{f14}
\end{figure}

An example of energy exchange between conjugate high-frequency modes, which gives rise to the energy transfer along the chain is shown in the figures \ref{f13}-\ref{f14}.

\subsection*{Spatial energy localization in the discrete FPU chains.}

The dynamics of the normal modes in the essentially discrete system considered above has been studied in the assumption that the gap between $\pi $-mode and nearby conjugate ones is large enough. But, if the number of particles grows the new opportunity arises. Namely, the resonant interaction between the modes having close frequencies (in linear approximation limit) becomes possible. Really, the frequency of $\pi $-mode differs from the frequency of nearby ones by the value of order $1/N^2$:

\begin{equation} \label{eq43} 
\omega _{\frac{N}{2} \pm 1} =2\sin (\frac{\pi }{N} (\frac{N}{2} \pm 1))=2\cos (\frac{\pi }{N} )\approx \omega _{\frac{N}{2} } (1-\frac{1}{2} (\frac{\pi }{N} )^{2} ).  
\end{equation}

Because the corresponding integral manifolds, two qualitatively different scenarios can be single out. For small enough energies a simple superposition of the normal vibrations occurs that is similar to linear approximation. However, after attaining certain threshold the resonant interaction mentioned above occurs and it leads to some localization of excitation which can be move along the chain. This change in the behavior of the system can be interpreted as a manifestation of instability of invariant manifold in space of normal mode. The instability of zone-boundary $\pi $-mode, well known in the FPU-chains, usually corresponds with onset of weak chaos and the generation of chaotic breathers ~[\cite{book12}]. However, the results of the previous consideration show that the equation of dynamics of $\pi $-mode can be effectively linearized. So nothing instability of $\pi $-mode has been observed. Even if an initial excitation is distributed between several modes (the $\pi $-mode and nearby ones) the resulting dynamics of the chain is well described by simple superposition of these modes. The computer simulation data are in good accordance with the analytical model.

While the energy of excitation and/or number of particles grows the dynamics of the chain becomes different than that for the quasi-harmonic dynamics defined by equations (\ref{eq18}). While the parameter $\varepsilon $ decreases, the eigenvalues of the $\pi $-mode and the nearby ones become more and more close until the resonant conditions occur. 

In such a case the equations (\ref{eq33}) are not valid because this resonance does not take into account. Now we choose the top frequency mode $\chi_{N/2}$ as the basic one. So, because the difference between $\chi_{N/2}$ and $\chi_{N/2-1}$ has the order $\varepsilon^2$ additional linear items appear in the equations for $\chi_{N/2-1}$ and $\chi_{N/2+1}$. Moreover, we have to take into account the resonant interactions in the nonlinear part of equations. Missing out the details of calculations we get the final equations in the form:

\begin{equation} \label{eq44} 
\begin{array}{l} {i\frac{d\chi _{N/2} }{d\tau _{2} } +\frac{3\beta }{4} [|\chi _{N/2} |^{2} \chi _{N/2} +2(|\chi _{N/2-1} |^{2} +|\chi _{N/2+1} |^{2} )\chi _{N/2} +} \\ {\quad \quad \quad \quad \quad \quad \quad \quad \quad \quad \quad \quad (\chi _{N/2-1}^{2} +\chi _{N/2+1}^{2} )\chi _{N/2}^{*} ]=0} \\ {i\frac{d\chi _{N/2-1} }{d\tau _{2} } -\frac{\pi ^{2} }{2} \chi _{N/2-1} +\frac{3\beta }{8} [(4|\chi _{N/2} |^{2} +3|\chi _{N/2-1} |^{2} +} \\ {\quad \quad \quad \quad \quad 2|\chi _{N/2+1} |^{2} )\chi _{N/2-1} +(2\chi _{N/2}^{2} +\chi _{N/2+1}^{2} )\chi _{N/2-1}^{*} ]=0} \\ {i\frac{d\chi _{N/2+1} }{d\tau _{2} } -\frac{\pi ^{2} }{2} \chi _{N/2+1} +\frac{3\beta }{8} [(4|\chi _{N/2} |^{2} +3|\chi _{N/2+1} |^{2} +} \\ {\quad \quad \quad \quad \quad 2|\chi _{N/2-1} |^{2} )\chi _{N/2+1} +(2\chi _{N/2}^{2} +\chi _{N/2-1}^{2} )\chi _{N/2+1}^{*} ]=0} \end{array}.  
\end{equation} 

First of all one should point out the presence of ``exchange'' items $ (\chi_{j}^2\chi_{N/2}^{*}$, $ j=N/2+1,N/2-1 ) $ in the equations (\ref{eq44}) which are analogous ones to those in the equations for conjugate modes (\ref{eq33}). It is easy to check that equations (\ref{eq44}) possess an additional integral of motion -- total ``occupation number''

\[X=|\chi _{0} |^{2} +|\chi _{1} |^{2} +|\chi _{2} |^{2} =const. \] 

\begin{figure}[htbp]
            \centering
       \includegraphics[width=49.9mm, height=39.6mm]{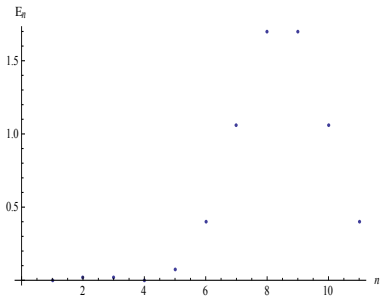}
        \caption{The energy distribution in the $\beta $-FPU system of 12 particles corresponding to the excitation of $\pi $-mode in combination with one of conjugate mode.}
	\label{f15}
    \end{figure}

One can see that the equations (\ref{eq44}) admit both the one-mode and the two-mode solutions.  If $\chi_{N/2}=0$ is the choice we get the equations that are analogous of equations (\ref{eq33}), because the presence of the linear items does not introduce any significant distinctions. The computer simulation data confirm the existence of dynamical regime with only excited conjugate modes even in the long chains. On the other side, it is easy to show that a combination of modes with different eigenvalues (i.e. a combination of $\pi $-mode and one of conjugate mode) leads to the localization of energy in a one half of the chain (see the fig. \ref{f15}).

And at last, the energy put only in the mode with specified ``parity'' (e.g. $\chi_{N/2-1}$) is not transferred into mode with other ``parity'' (one can show that this statement is destructed by instability of conjugate modes). Let us consider the equations (\ref{eq44}) with initial conditions ($\chi_{N/2+1}=0$, $\chi_{N/2}\not=0$, $\chi_{N/2-1}\not=0$) for the determinacy. In such a case to analyze the dynamics of the system it is convenient to introduce new variables $\varphi_0$, $\varphi_1$:

\begin{equation} \label{eq45} 
\begin{array}{l} {\chi _{N/2} =\frac{1}{\sqrt{2} } (\varphi _{0} +\varphi _{1} )} \\ {\chi _{N/2-1} =\frac{1}{\sqrt{2} } (\varphi _{0} -\varphi _{1} )} \end{array}.  
\end{equation} 

Because the ``occupation number'' $X=|\varphi_0|^2+|\varphi_1|^2$ is the integral of motion as before, the variables $\varphi_0$ and $\varphi_1$ can be expressed by the angle variables:

\begin{equation} \label{eq46} 
\begin{array}{l} {\varphi _{0} =\sqrt{X} \cos \theta e^{i\delta _{0} } } \\ {\varphi _{1} =\sqrt{X} \sin \theta e^{i\delta _{1} } } \end{array}.  
\end{equation} 

The Hamilton function, written in the terms ($\theta $, $\Delta=\delta_1-\delta_0$), has the form:

\begin{equation} \label{eq47} 
\begin{array}{l} {H(\theta ,\Delta )=\frac{X}{64} [51\beta X-16\pi ^{2} +2(8\pi ^{2} -3\beta X)\cos \Delta \sin 2\theta } \\ {-3\beta X(8-\cos ^{2} \Delta )\sin ^{2} 2\theta ]} \end{array}.  
\end{equation} 

The respective equations of motion 

\begin{equation} \label{eq48} 
\begin{array}{l} {\frac{\partial \theta }{\partial \tau _{2} } +\frac{1}{32} (8\pi ^{2} -3\beta X+3\beta X\cos \Delta \sin 2\theta )\sin \Delta =0} \\ {\sin 2\theta \frac{\partial \Delta }{\partial \tau _{2} } -\frac{1}{32} \cos 2\theta [(8\pi ^{2} -3\beta X)\cos \Delta -3\beta X(8-\cos ^{2} \Delta )\sin 2\theta ]=0} \\ {\Delta =\delta _{1} -\delta _{0} } \end{array} 
\end{equation} 

have the stationary points:

\begin{equation} \label{eq49} 
\begin{array}{l} {(a)\quad \Delta =0,\quad \theta =\frac{\pi }{4} } \\ {(b)\quad \Delta =\pi ,\quad \theta =\frac{\pi }{4} } \\ {(c)\quad \Delta =0,\quad \theta _{1} =\frac{1}{2} \arcsin (\frac{8\pi ^{2} -3\beta X}{21\beta X} ),\quad \theta _{2} =\frac{\pi }{2} -\theta _{1} } \end{array}.  
\end{equation} 

Two first stationary points ((a) and (b)) correspond to equal values of $|\varphi_0|$  and $|\varphi_1|$ , and it leads to the pure $\pi $-mode (a) or pure $\chi_{N/2-1}$-mode (b) in the accordance with definition (\ref{eq45}). These stationary states exist at every values of the amplitude $X$. The limiting phase trajectory (LPT) rounding the stationary point (a) corresponds to the transfer from the state where $\chi_{N/2}=\chi_{N/2-1}$ to the state where $\chi_{N/2}=-\chi_{N/2-1}$, that leads to energy transition from one half of the chain to another one. That is the nonlinear analogue of beating phenomena in the linear system. In the small-amplitude limit it is the only mechanism of the energy redistribution in the chain, when the energy exchange between conjugate modes is absent.

\begin{figure}[htbp]
            \centering
        (a) \includegraphics[width=54.4mm, height=52.1mm]{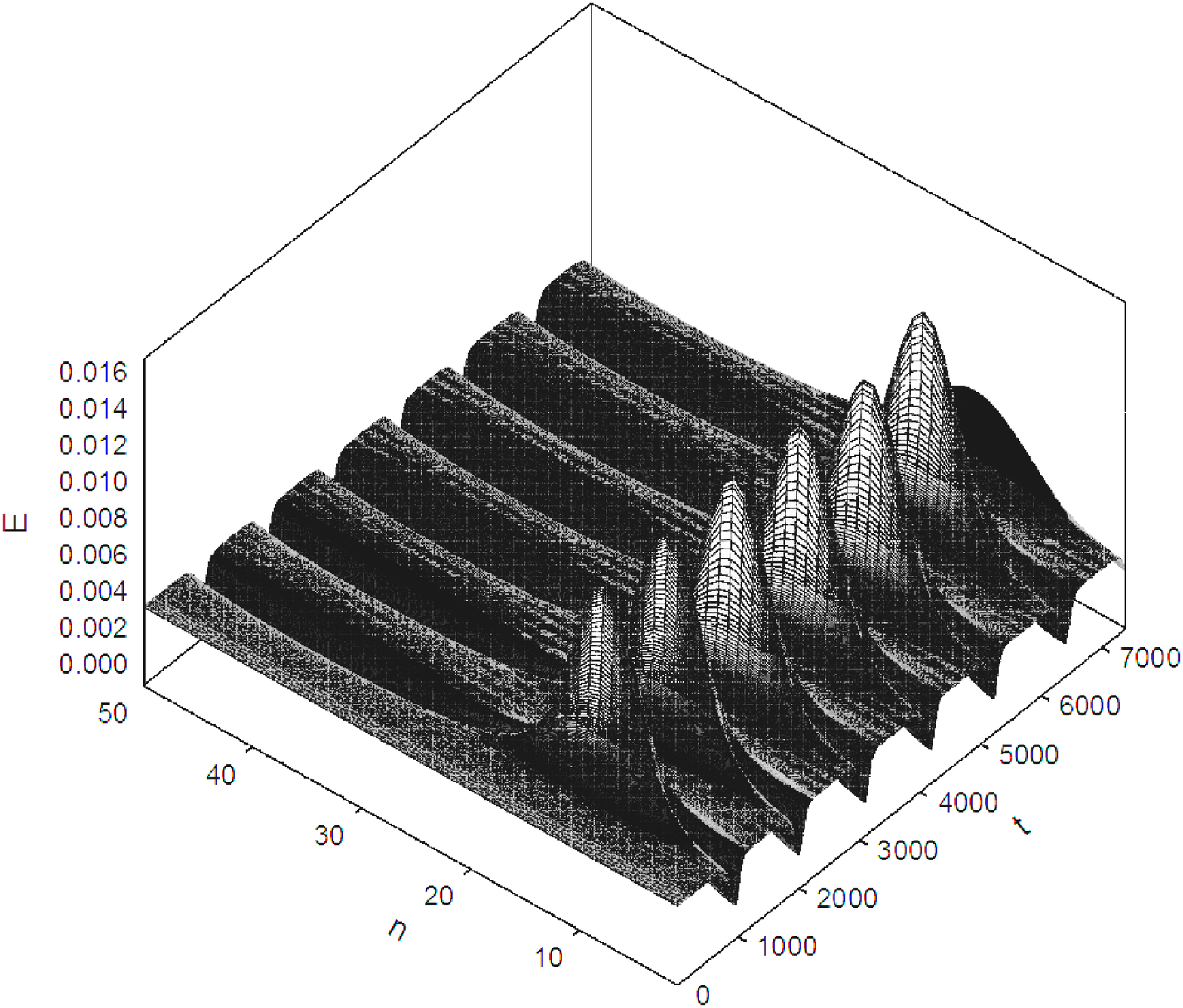}(b) \includegraphics[width=54.4mm, height=52.1mm]{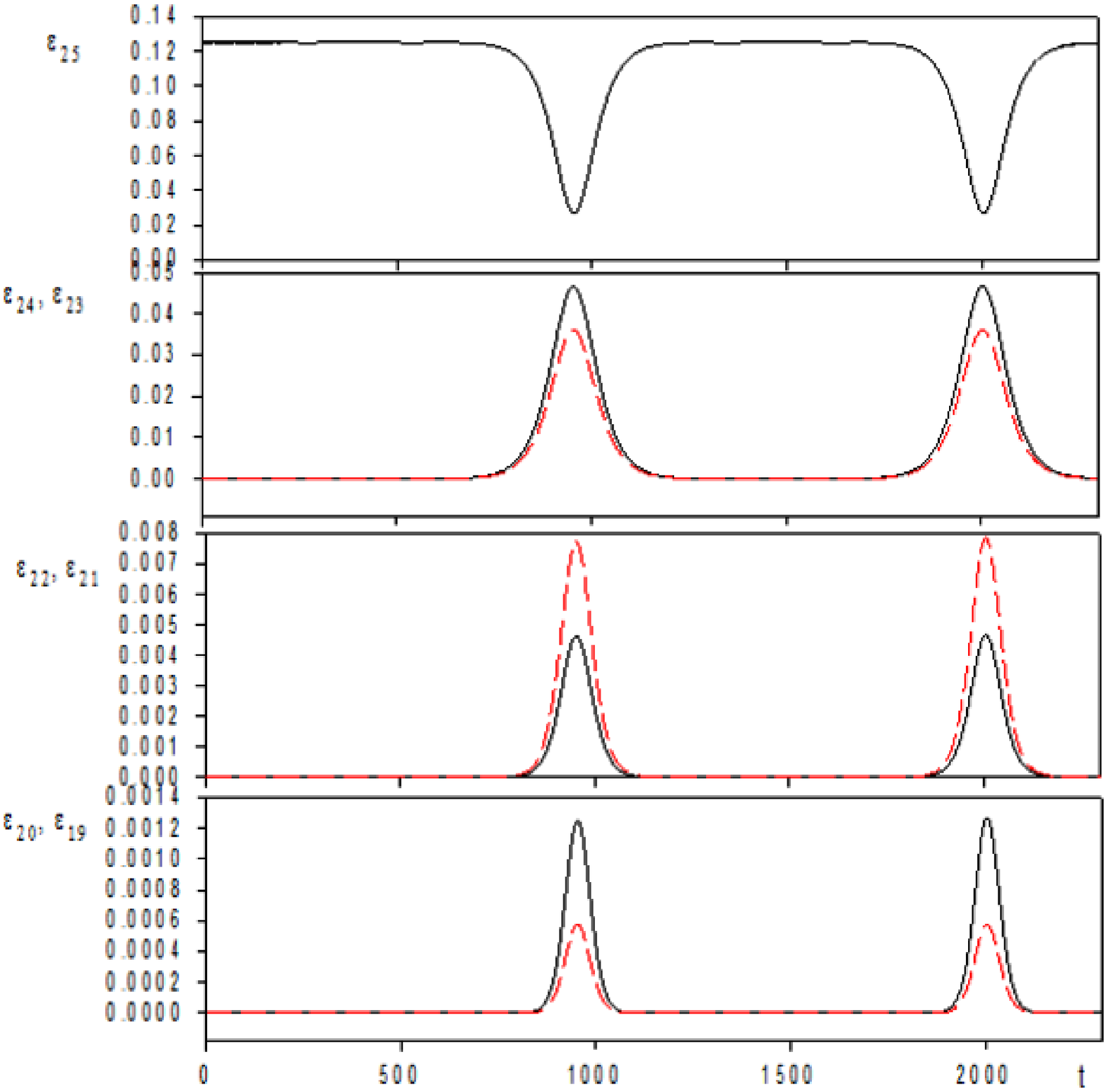}
        \caption{Instability of $\pi $-mode in the $\beta $-FPU chain with 50 particles: (a) 3D surface of energy of particles; (b) energy of normal modes. $E$ -- energy of particle, $n$ -- number of particle, $t$ -- time, $\varepsilon_k$ -- energy of normal modes; the initial occupation number $X_0=6.25$.
}
	\label{f16}
    \end{figure}

The stationary points (c) originate from the point (a) when the amplitude $X$ reaches the value $\pi^2/3\beta $. At this instant the stationary point (a) becomes the saddle one, consequently the zone-boundary $\pi $-mode loses its stability. The instability of $\pi $-mode appears as a small admixture of nearby conjugate mode that leads to a weak modulation of amplitude of original mode. If the amplitude $X$ reaches large enough values, the modulation of $\pi $-mode turns out be large, that leads to generation of short life-time breathers. The example of such instability in the chain with 50 particles is shown in the fig. \ref{f16}. The energy of the chain corresponding to the threshold value X is equal to $\pi^2/3\beta N$ (only own part of the mode energy is taken into account). This result is in the good accordance with the estimation ~[\cite{book12}] in the limit of the large chain ($N \gg 1$).

\begin{figure}[htbp]
            \centering
        (a) \includegraphics[width=40.1mm, height=40.2mm]{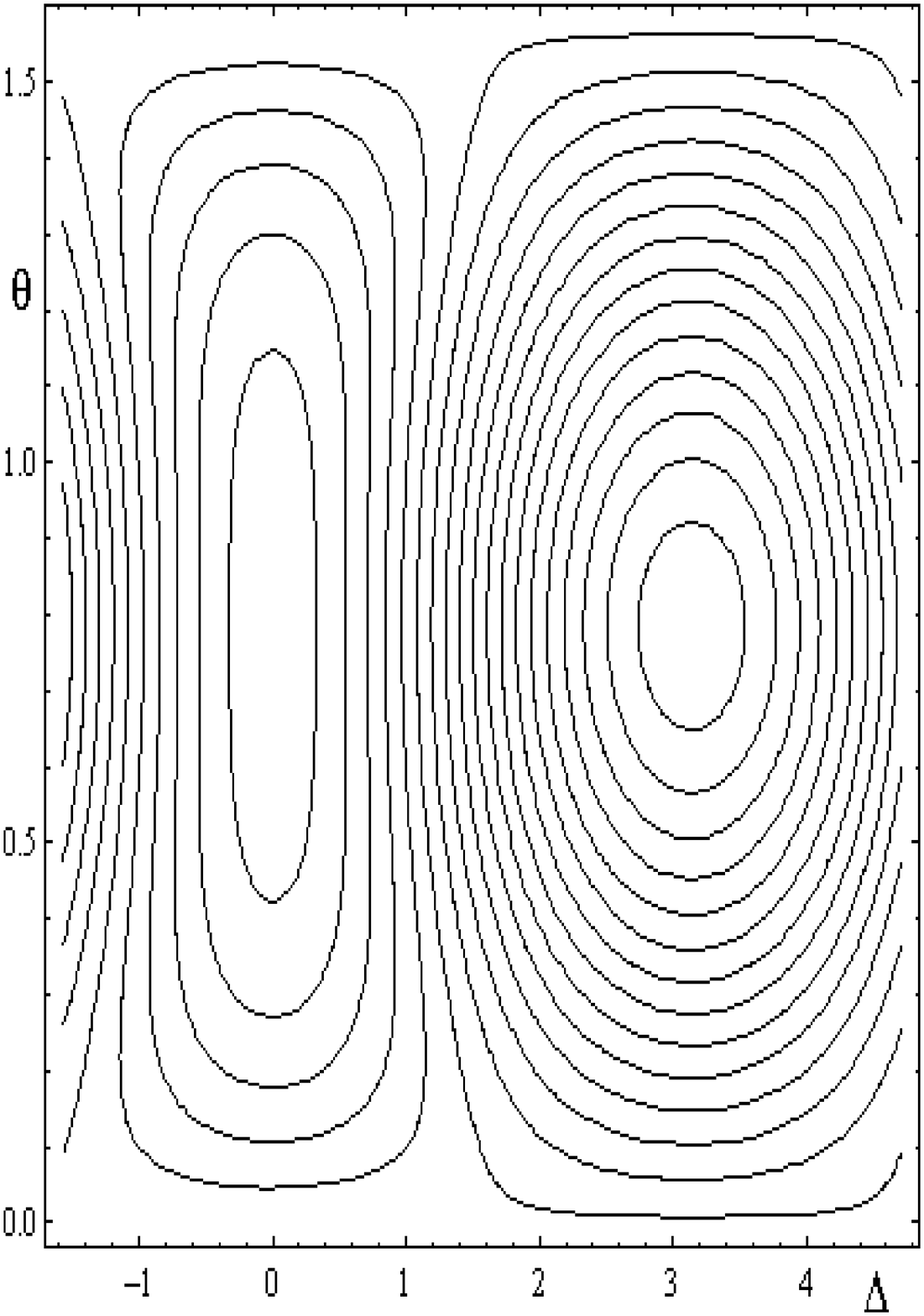}(b) \includegraphics[width=40.1mm, height=40.2mm]{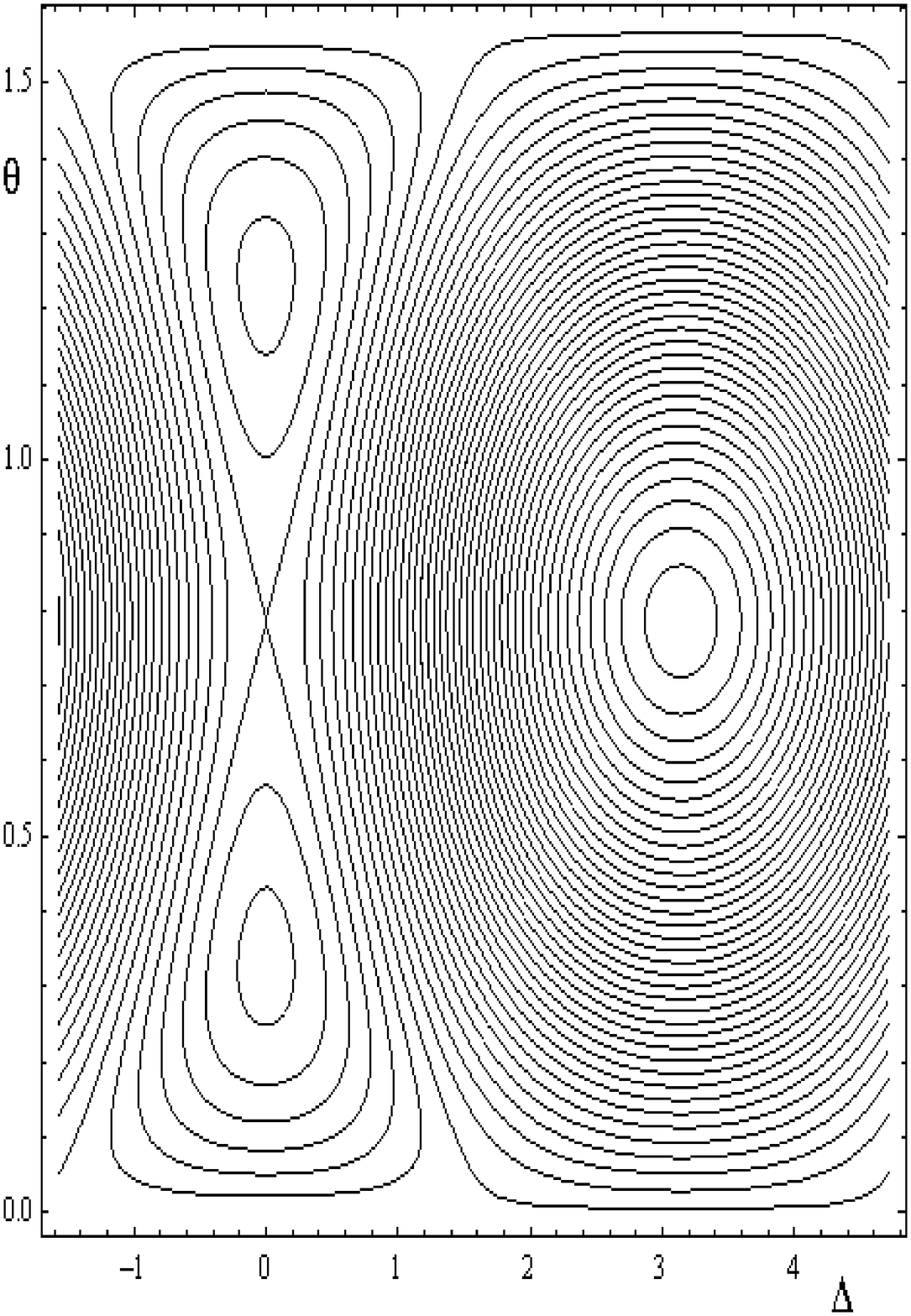}

        (c) \includegraphics[width=40.1mm, height=40.2mm]{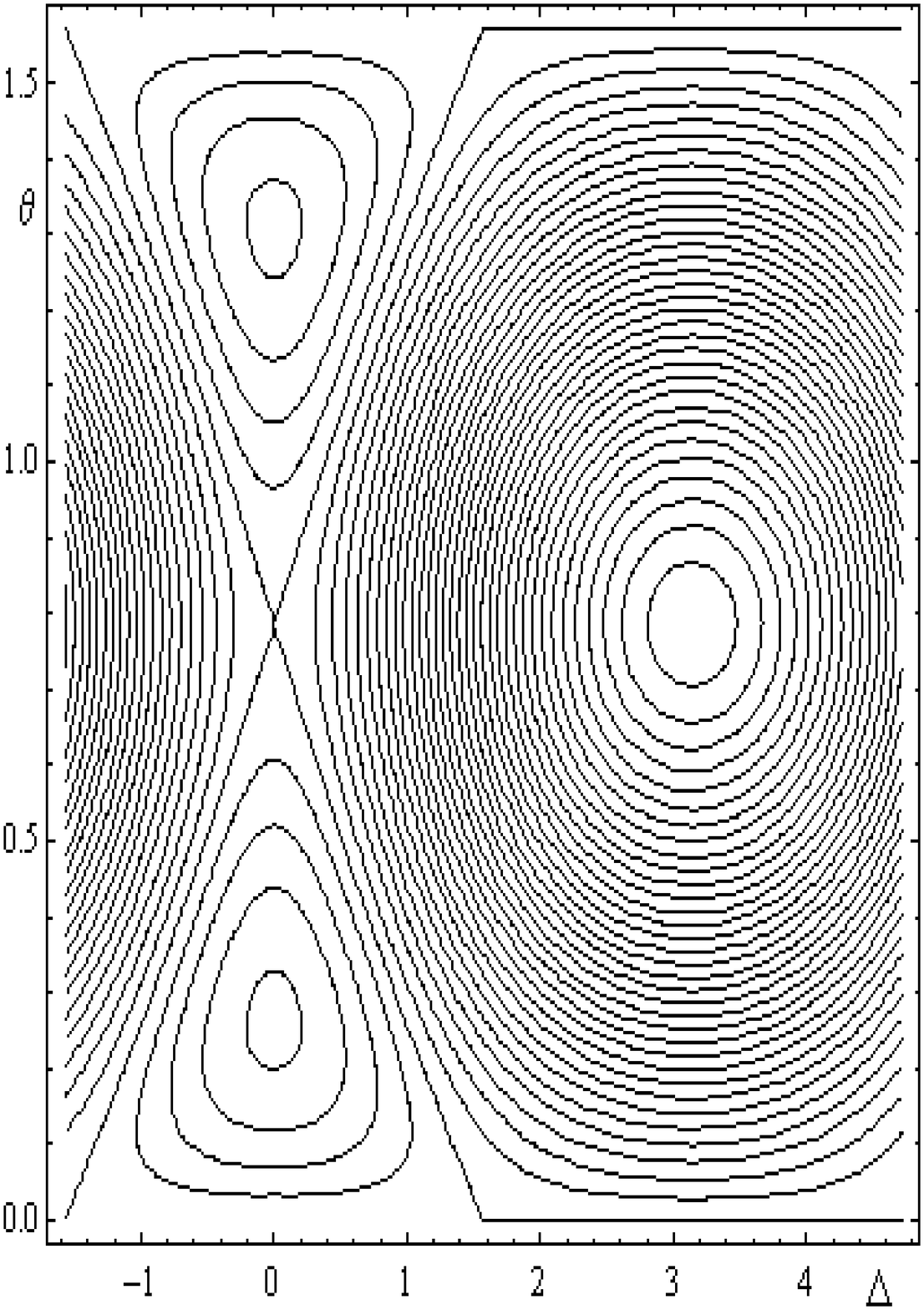}(d) \includegraphics[width=40.1mm, height=40.2mm]{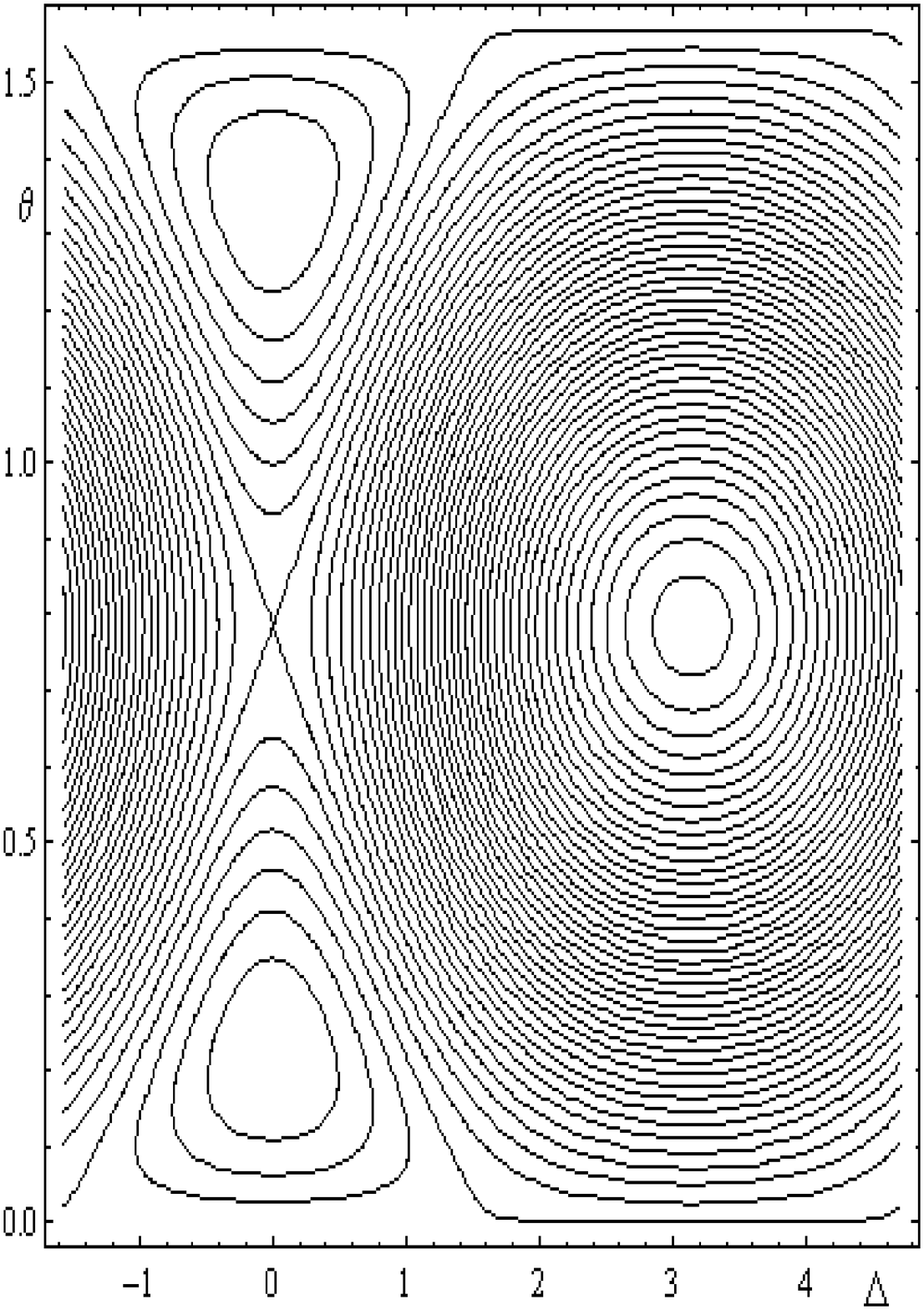}
        \caption{Phase plane transformation when the amplitude X grows. (a)$\beta X=\pi^2/3$ Phase plane before the first bifurcation: there are only two singular points like center; (b) $\beta X=\pi^2/2$ after the first bifurcation the separatrix crossing the stationary point ($0, \pi/4$) describes the instability of $\pi $-mode; (c) the LPT merges with the separatrix at the critical value of occupation number $X_c=16\pi^2/27\beta $, the transitions from any state below $\theta =\pi /4$ near $\theta =0$ to the state with $\theta>\pi/4$ are forbidden and vice versa; (d) $\beta X=2\pi^2/3$ - the area of transit-time trajectories is observed at the any occupation number $X>X_c$ and LPT is limited by the amplitude, corresponding to the angle $\theta <\pi/4$.}
	\label{f17}
    \end{figure}

While the amplitude $X$ grows, the separatrix rounding the stationary points (c) enlarges but the transition along the LPT is possible up to the instant when the separatrix merges with the LPT. Since this instant the trajectories starting from the point with $\theta < \pi/4$ and $\theta =0$ cannot reach any point with $\theta>\pi /4$. It means that the energy of initial excitations turns out to be confined at the one part of the chain. The threshold of such localization can be calculated from the requirement of the equality of energy of the states in the LPT and the energy of unstable stationary point (a). The result is $X_c=16\pi^2/27\beta $ that corresponds to the energy threshold $E_loc=16\pi^2/27\beta N$. It is convenient to observe the evolution of phase trajectories accompanied the variation of amplitude $X$ on the phase space in the terms of angle variables $\Delta $ and $\theta $. Four different structures of phase plane are shown in the fig. \ref{f17}. The first graph (\ref{f17}.a) shows the phase plane before the first bifurcation, i.e. at the small values of $X$. After the first bifurcation the small separatrix crossing the unstable point (0, $\pi /4$) originates at $X=\pi^2/3\beta $ (fig. \ref{f17}.b). The second bifurcation is shown in the fig. \ref{f17} (c). It is well seen that the LPT merges with the separatrix. The subsequent rise of amplitude $X$ leads to the creation of transit-time trajectory region (fig. \ref{f17} (d)).

The LPT starting the point $(0,~0)$ is the boundary trajectory divided the area of closed motion around the stationary point (0, $\theta_1$) from the area of transit-time trajectories. The new separatrix crossing the point (0, $\pi /4$) describes the transition ($\chi_{N/2} \rightarrow \chi_{N/2}$) and rounds the stationary point $\chi_{N/2-1}$. One should note that the LPT describes the periodical process which is accompanied by some deformation of energy distribution profile (the ``breathing'' mode of localized excitations -- see fig.\ref{f13}.b). The period of this ``breathing'' can be estimated as

\begin{equation} \label{eq50} 
T=\oint d\tau =\oint \frac{d\Delta }{d\Delta /d\tau }   ,  
\end{equation} 

where the last integral is taken along the LPT. The equation of LPT is defined by proposition

\[H(\theta ,\Delta )=H(0,0)=\frac{X}{64} (51\beta X-16\pi ^{2} ). \] 

This relationship leads to 

\begin{equation} \label{eq51} \begin{array}{l} {\sin 2\theta =2\kappa \frac{\cos \Delta }{8-\cos ^{2} \Delta } } \\ {\kappa =\frac{8\pi ^{2} -3\beta X}{3\beta X} } \end{array}.  \end{equation} 

\begin{figure}[htbp]
            \centering
        \includegraphics[width=67.9mm, height=40.3mm]{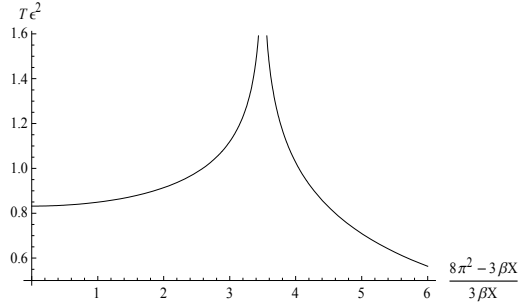}
        \caption{Period of motion along the LPT versus parameter $\kappa $. The value $ \kappa $ =3.5 corresponds to the threshold of localization $X_c=16\pi ^2/27\beta $. The smaller values of $ \kappa $ respect to the localized regime with $ X>X_c $.}
	\label{f18}
    \end{figure}

The variation of period as a function of parameter $\kappa $ is shown in the figure \ref{f18}.

\begin{figure}[htbp]
            \centering
        (a) \includegraphics[width=52.6mm, height=52.6mm]{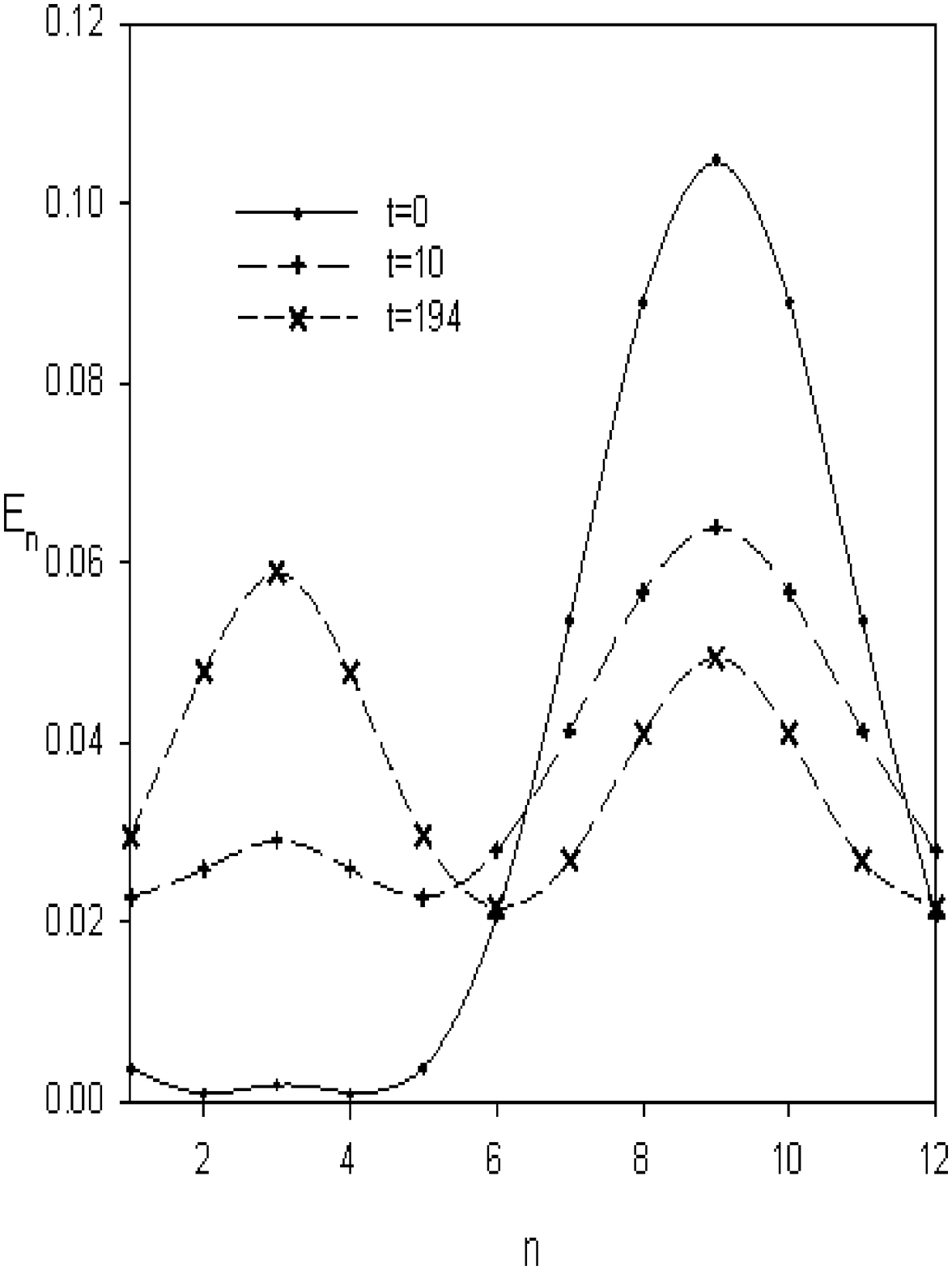} 
	(b) \includegraphics[width=52.6mm, height=52.6mm] {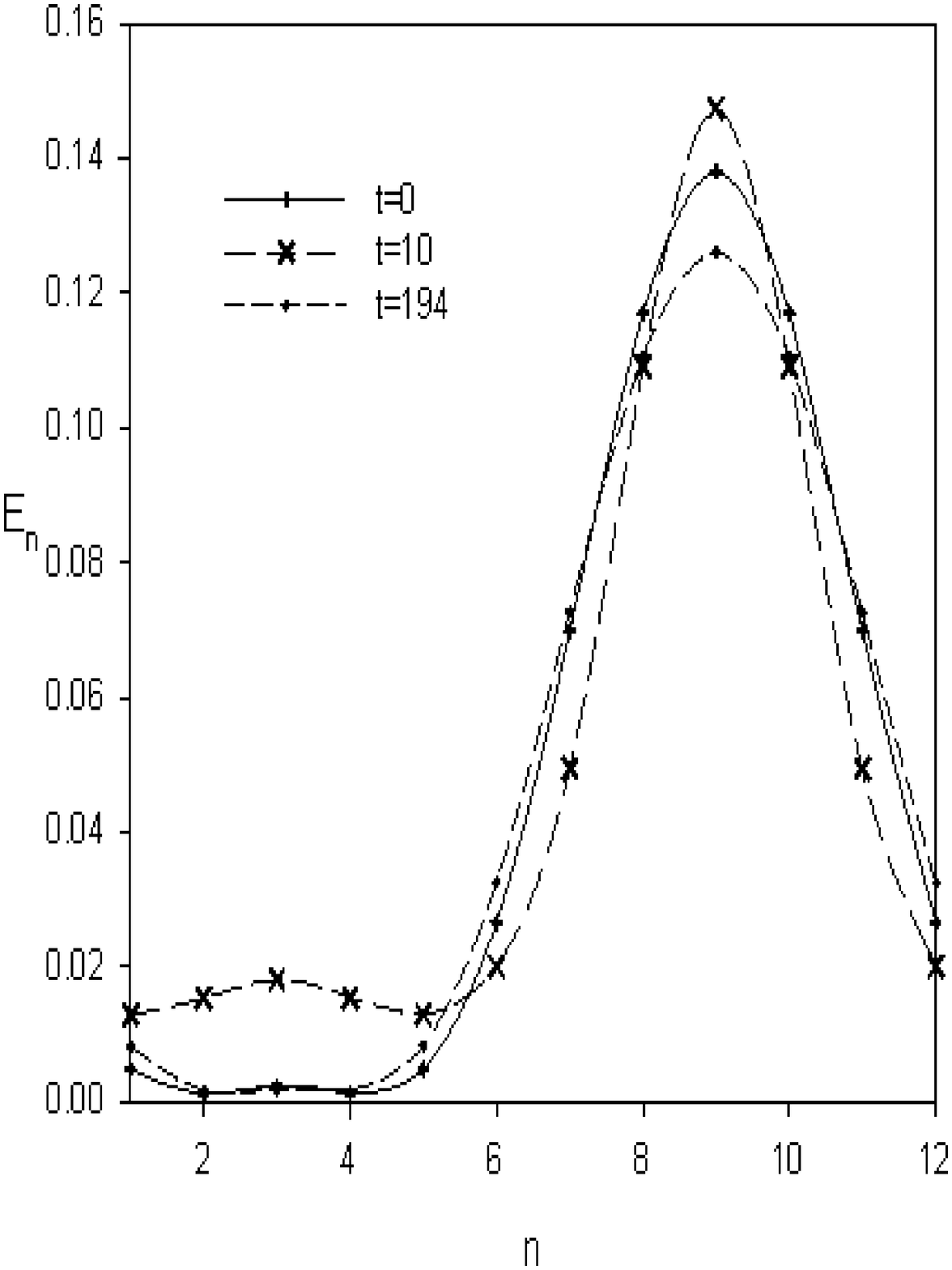}
        \caption{The energy distribution along the chain with 12 particles at $ X=14\pi^{2}/27 \beta <X_c$ (a) and $ X=2\pi^{2}/3 \beta >X_c $ (b) at the different times $ t $. The ``breathing'' mode of the energy profile in the right side corresponds to motion along the LPT (see text).}
	\label{f19}
    \end{figure}

In the fig. \ref{f19} the energy distribution along the chain with 12 particles is shown both before threshold of localization and after that.

\begin{figure}[htbp]
            \centering
	(a) \includegraphics[width=53.3mm, height=52.1mm]{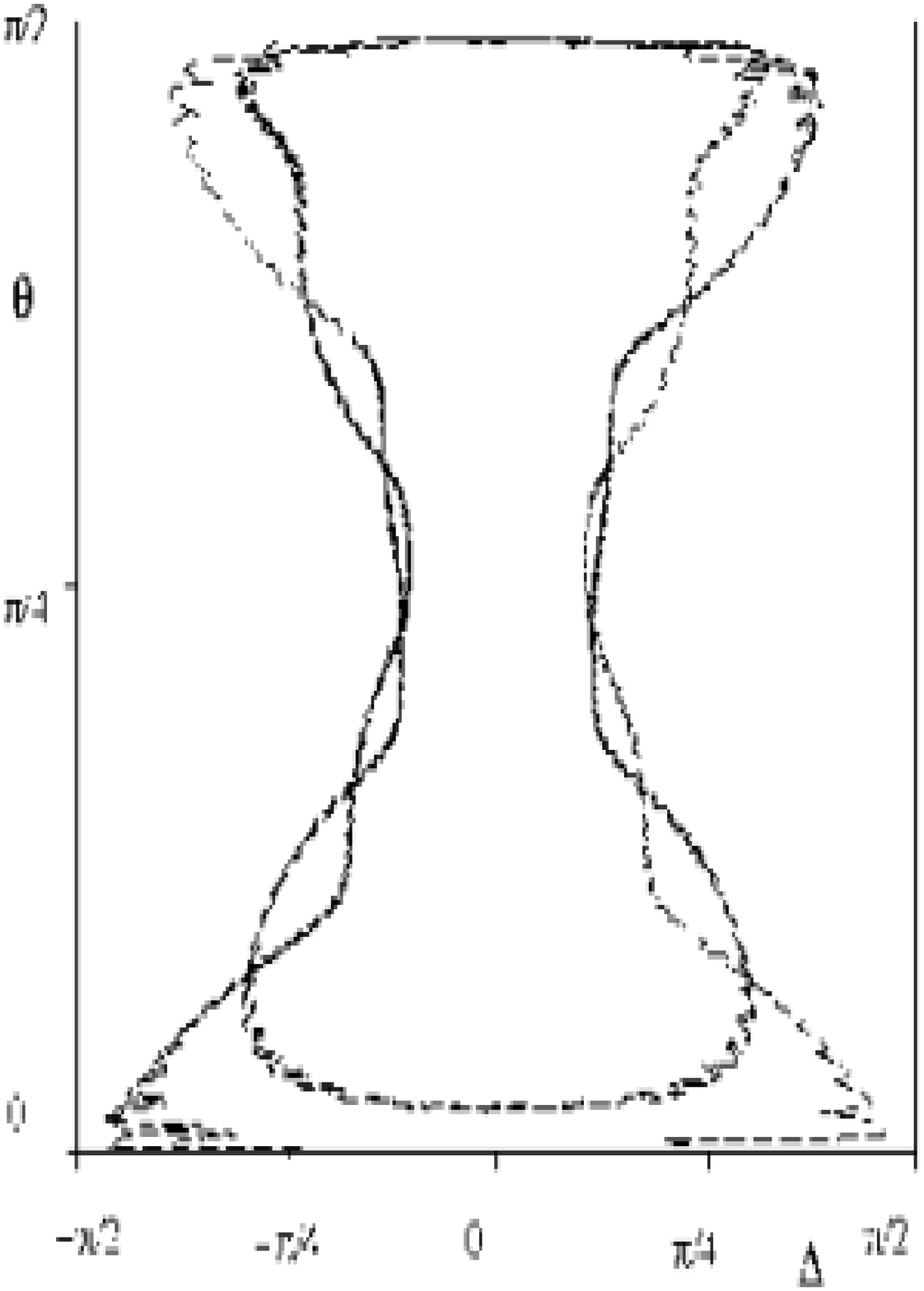} 
	(b) \includegraphics[width=53.3mm, height=52.1mm]{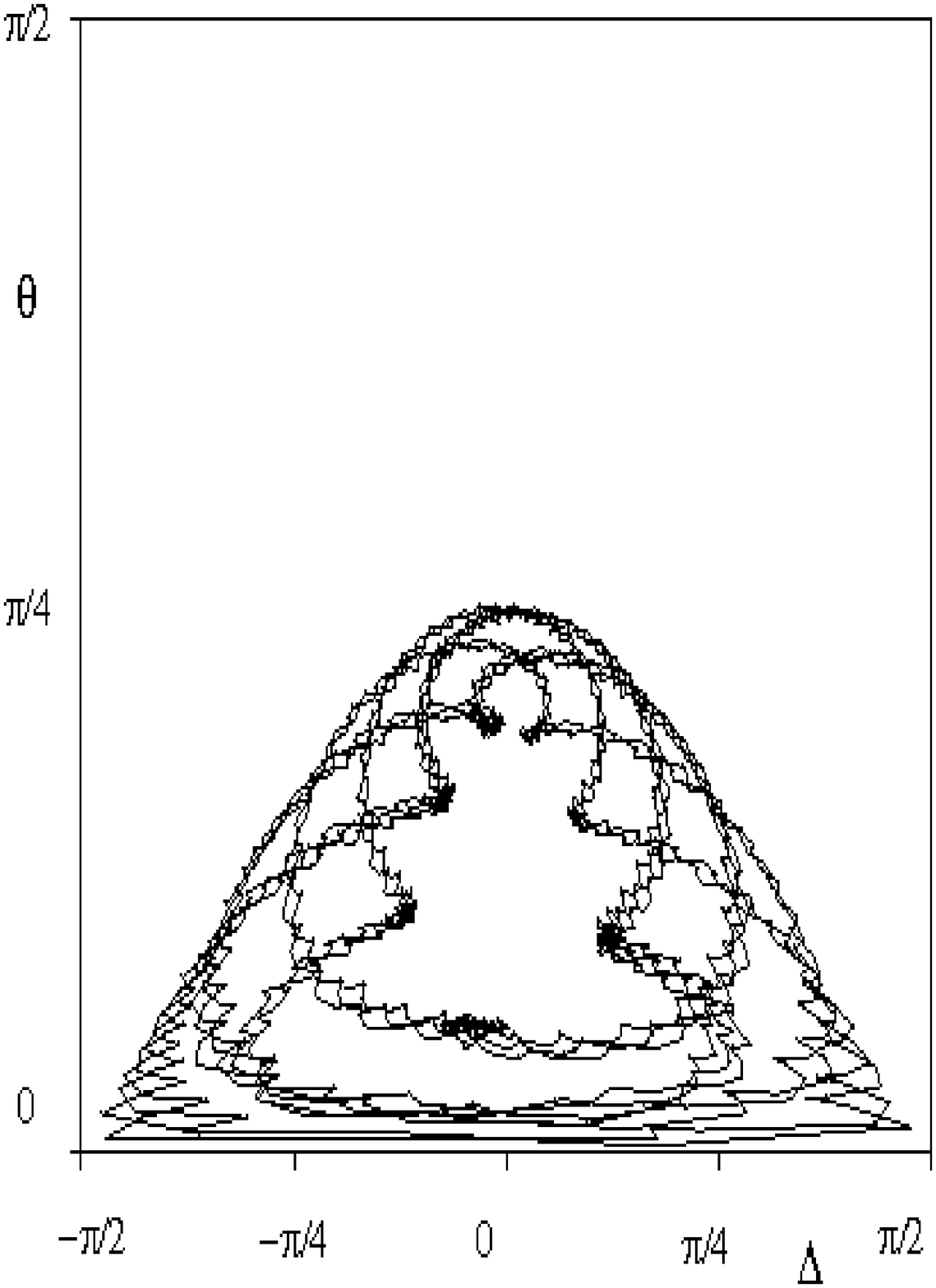}
        \caption{The LPT on the ($\theta, \Delta$)-plane for the chain with 12 particles: (a) --before the threshold of localization and (b) - after that. These trajectories correspond to the energy distributions showed in the fig. \ref{f19}.}
	\label{f20}
\end{figure}

So one can see that energy localization in the one half of chain needs a simple superposition of two modes with the total occupation number $X$ exceeding the critical value $X_c$. The value of $X_{c}$ does not depend on the length of the chain but the total energy E is directly proportional $\varepsilon X=X/N$ that leads to the infinitesimal threshold of localization in the case of infinite chains. Really this process is an outset of the creation of localized excitations (such as chaotic breathers) in the system. The computer simulation data show the energy localization in a more narrow area if the chain is the long enough ($ N\sim 20 $ and more). Such a contraction of localization area needs a participation of modes different from the considered ones. The presence of these additional modes is reflected in the phase portrait of the system (fig. \ref{f20}) where the computed phase trajectories are more complicated than the analytical predicted ones (fig. \ref{f17}). The figure 15 shows the long time localization in the system with 8 particles. One can see that the motion of the position of energy maximum correlates with energy exchange between conjugate modes $\chi_{N/2-1}$ and $\chi_{N/2+1}$. The reason of that is the difference between asymptotic equations (\ref{eq44}) and the exact dynamical behavior of the original FPU chain. These differences grow when the amplitude of excitations increase that manifests as an instability with respect to energy exchange between the conjugate modes considered. 

\begin{figure}[htbp]
            \centering
        (a)\includegraphics[width=58.5mm, height=59.6mm]{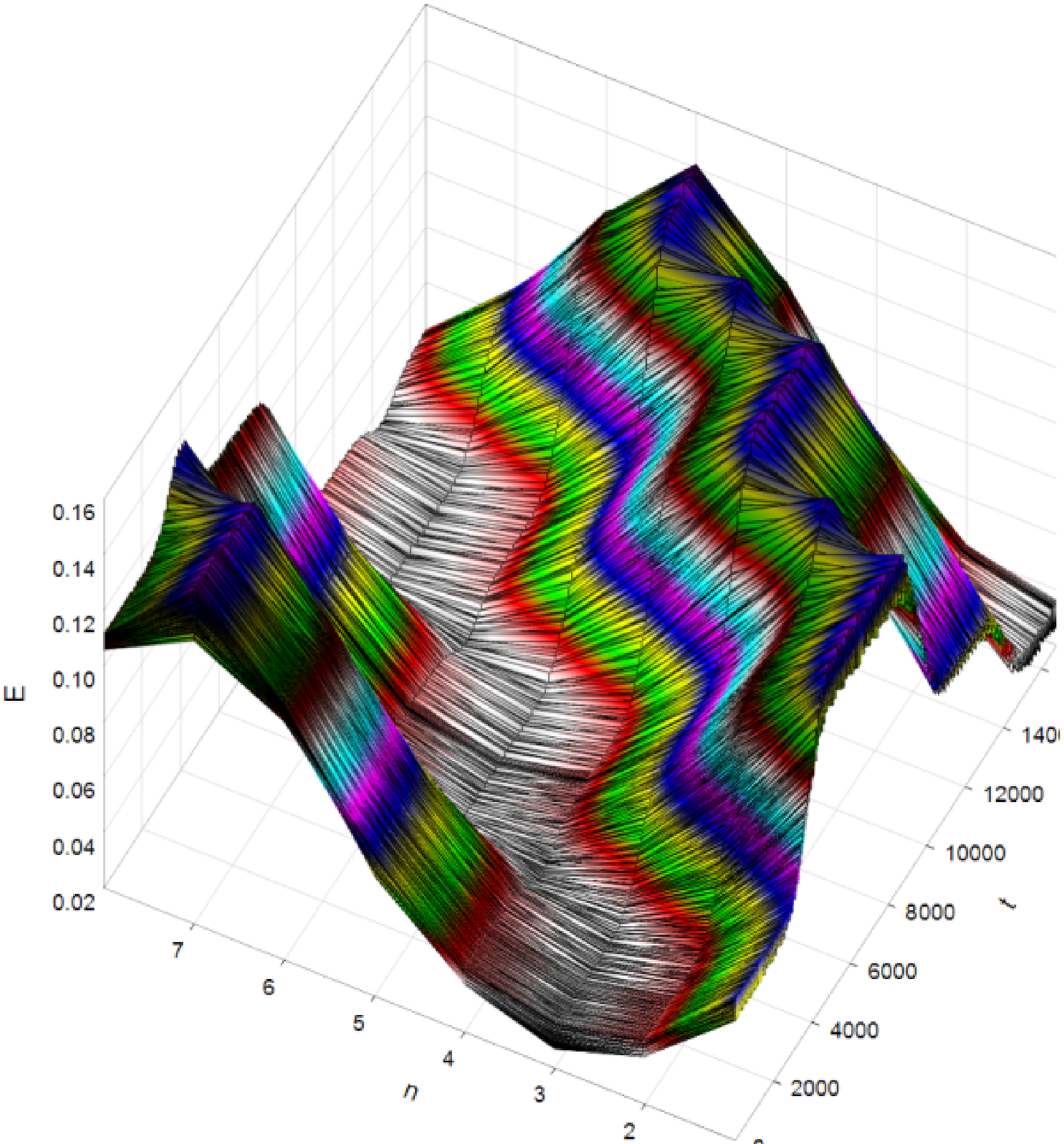} 
	(b)\includegraphics[width=58.5mm, height=59.6mm]{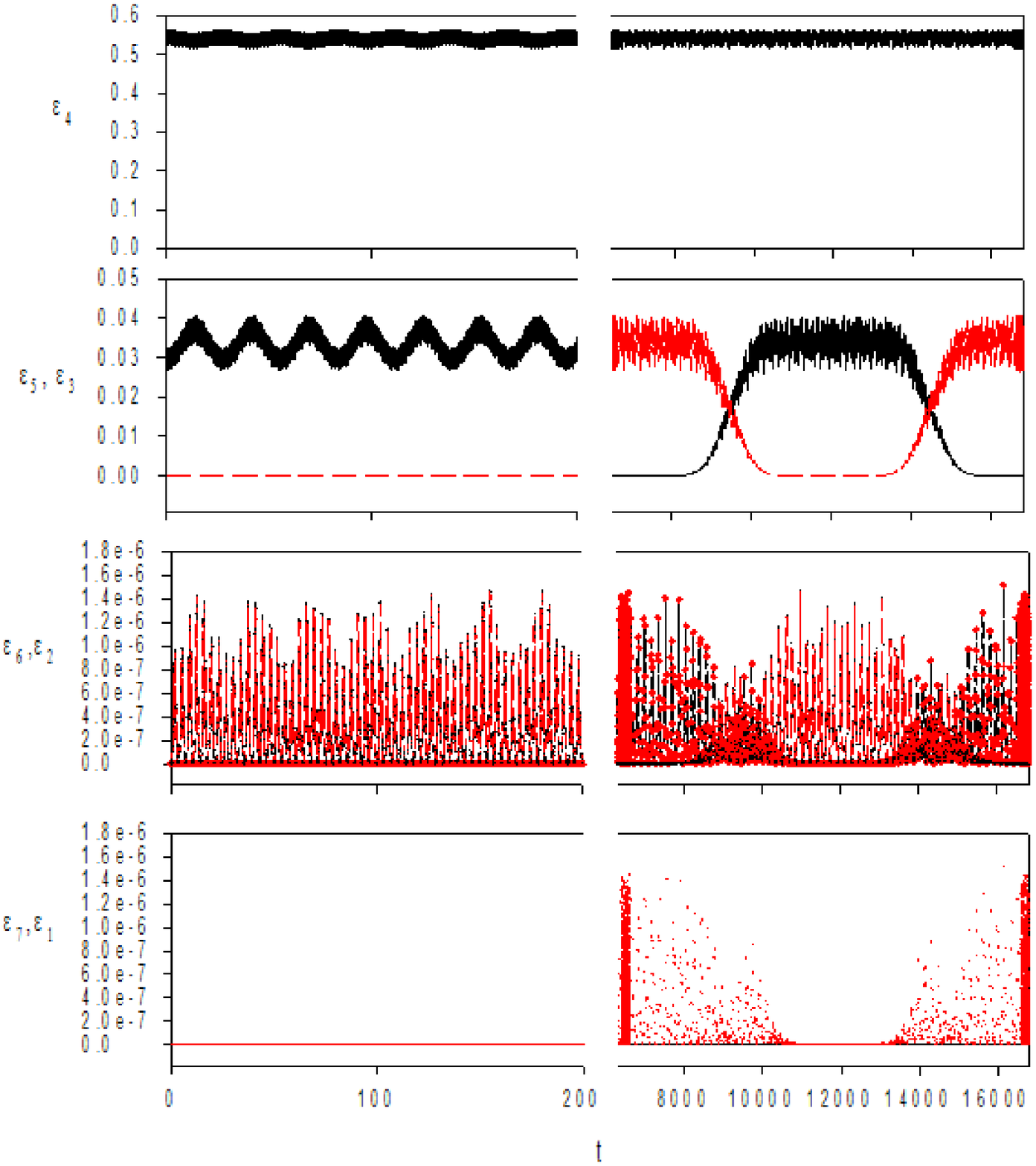}
        \caption{The computer simulation data for the chain consisting 8 particles: (a) The surface of energy of particles, (b) the energy of normal modes. The initial excitation corresponds to $ X=8.0$ .}
	\label{f21}
    \end{figure}

\subsection*{Conclusion.}

The dynamics of essentially discrete nonlinear systems with periodic boundary conditions demonstrates two typical regimes of behavior. The first of them is due to the discreteness of the spectrum of eigenvalues of the system and corresponds to the quasiperiodic motion along the invariant tori in the Fourier-space of the system. In such a case the dynamics of resonant pairs of conjugate modes in small FPU systems can be described from unique viewpoint using complex representation of dependent variables and multiple scale techniques. As this takes place, the complete general classification of possible types of dynamical behavior is attained. The most simple scenario is realized for the case of symmetric nonlinearity ($\beta$-FPU system) when only two characteristic phase trajectories corresponding to supernormal (elliptical) mode and LPT, can be discriminated. They describe the vibrations with phase shift $\Delta =0$ ($\Delta =\pi/2$) between two resonant modes and complete energy transfer between them, respectively. Both intensity and rate of intermode energy exchange are controlled (depending on the resonant pair) by parameters $G_{1}^{+}/X_1$ or $G_{1}^{-}/X_1$ (despite the difference between $G_{1}^{+}/X_1$ and $G_{1}^{-}/X_1$, corresponding scenario are similar except of the presence of supernormal mode in the first case and elliptical mode in the second one). Rise of this parameters leads to decreasing of exchange intensity and increasing of its rate. It is very important that the evolution of initial excitation is described efficiently by the set of quasilinear differential equations parametrized by the ``occupation number'' of the respective normal modes $X$.

Introduction of asymmetry leads to complication of the dynamics of the conjugate modes with indexes $N/4$ and $3N/4$ with several qualitative transitions. The evolution of dynamical behavior with increasing of asymmetry is inverse (in certain sense) to that for linearly coupled nonlinear oscillators. For strong asymmetry two stable stationary states corresponding to supernormal and elliptical modes exist. Two branches of LPT describe complete energy exchange. When parameter of asymmetry decreases the instability of supernormal mode occurs with formation of singular trajectory (separatrix) and breaking off the complete energy transfer. Further evolution with decreasing of the parameter of asymmetry leads to full termination of energy exchange between resonant modes, restoring the stability of supernormal mode with simultaneous instability of elliptical one and possibility of partial intermode energy exchange. If asymmetry disappears the separatrix transform to LPT with complete energy transfer. 

The second dynamic regime is associated with the interaction between non-conjugate high-frequency modes, i.e. modes with different eigenvalues in the linear limiting case. However, the small splitting of the frequencies of zone-boundary $\pi$-mode and modes nearby of it provides almost resonant conditions between them. This circumstance can be the consequence of both elongation of chain length and the rise of the amplitude of oscillations. It is very interesting that the natural small parameter here is the inverse length of the chain 1/N. The quasi-resonant interaction is a reason of instability of $\pi $-mode when the energy of the chain reaches the first threshold value $E=\pi^2/3\beta N$. This energy corresponds to the occupation number $X=\pi^2/3\beta $, which does not depend on the length of the chain. The subsequent rise of occupation number up to value $X_c=16\pi^2/27\beta $ leads to the confinement of the energy in the local area of the chain. The reason is that the separatrix crossing the unstable stationary point corresponding the $\pi $-mode increases and merges with LPT. This circumstance leads to the breaking of trajectories corresponding to the energy transfer from one half of the chain to other one. The localized excitation corresponds to new LPT demarcates the areas of closed trajectories and transit-time one. The energy profile of localized excitation undergoes some disturbances while the imaging point of the system moves along the LPT. It is a cause of ``breathing'' mode of the chaotic breathers a starting point of which is the process of localization discussed above. Moreover increasing of the amplitude of oscillations in the nonlinear system leads to the frequency shift that makes the interaction of high-frequency modes with more low frequency ones. The computer simulation data show some contraction of area of energy localization accompanied by the excitation of low-frequency modes. So the chaotic breathers result to the interaction of many modes with different eigenvalues regardless of the fact that the contribution of low-frequency modes can be small enough. The last remark concerns with the role of boundary conditions. In spite of the fact that the periodicity of the system studied is very essential proposition the systems with other boundary conditions can be considered in the framework of this approach. The decreasing of frequency splitting of the top modes with different eigenvalues occurs while both the length of the chain and the amplitude of oscillations increase. Therefore the efficient resonant interaction can result in the localization and confinement of energy in the bounded area of the chain.

Acknowledgements

The work was supported by Program of Department of Chemistry and Material Science (Program \#1), Russia Academy of Sciences, and Russia Basic Research  Foundation (grant  08-03-00420-à).

%
%
\par       
%

\end{document}